\documentclass{jfm}

\usepackage{graphicx}
\usepackage{natbib}
\usepackage{hyperref}
\usepackage{epstopdf, epsfig,amsmath}
\usepackage[percent]{overpic}
\usepackage{subfigure}
\usepackage{array}
\usepackage{xcolor}
\usepackage{tikz}
\usepackage{caption}
\usepackage{color}

\hypersetup{
    colorlinks = true,
    urlcolor   = blue,
    citecolor  = black,
}

\def\r#1{(\ref{#1})}
\def\aaa{\textit{a}}
\def\bbb{\textit{b}}
\def\ccc{\textit{c}}

\definecolor{purple0}{rgb}{0.63,0.13,0.94}
\definecolor{c00}{rgb}{0,1,1}
\definecolor{m00}{rgb}{1,0,1}

\title{Interaction between near-wall streaks and large-scale motions in turbulent channel flows}

\author{Zisong Zhou\aff{1},
  Chun-Xiao Xu\aff{1}
  \corresp{\email{xucx@mail.tsinghua.edu.cn}},
  \and Javier Jim{\'e}nez\aff{2}}

\affiliation{\aff{1}AML, Department of Engineering Mechanics, Tsinghua University, Beijing 100084, P.R. China
\aff{2}School of Aeronautics, Universidad Polit{\'e}cnica de Madrid, 28040 Madrid, Spain}

\begin{document}

\maketitle

\begin{abstract}

The interactions between the near-wall streaks and the large-scale motions (LSMs) of the
outer region of wall-bounded turbulent flows are investigated. The co-supporting hypothesis
of \citet{toh2005interaction} is checked in full-scale channels at low to moderate Reynolds
numbers, from two points of view. To study the top-down influence of the outer structures on
the spanwise motion of the near-wall streaks, a method inspired by particle-image
velocimetry is used to track the spanwise position of the streaks. Their spanwise advection
velocity is found to be affected by the hierarchy of large-scale circulations in the
logarithmic layer, but their spanwise streak density is only weakly related to the LSMs. The
evidence suggests that a top-down influence exists and drives the drift of the streaks in
the spanwise direction, as suggested by \citet{toh2005interaction}, but that the
hypothesized streak accumulation rarely occurs. Numerical experiments at
$\Rey_{\tau}\thickapprox535$ are then performed to clarify the role of the near-wall streaks
in the generation and preservation of the outer LSMs. The results show that the merger of
the near-wall streaks is only weakly correlated with the generation of the LSMs, and that
removing the near-wall roots of the LSMs does not affect the evolution of their outer
region. It is concluded that the bottom-up influence from the near-wall streaks is not
essential for the LSM generation and preservation, also weakening the evidence for the
co-supporting hypothesis.

\end{abstract}

\begin{keywords}
\end{keywords}

\section{Introduction}\label{sec:intro}

Coherent structures in wall-bounded turbulent flows play vital roles in turbulence
production and maintenance. Velocity streaks and quasi-streamwise vortices are the two
dominant structures in the near-wall region. Their cyclic generation consists of a
self-sustaining process \citep{jimenez1991minimal,hamilton1995regeneration}, which can
survive in the absence of turbulence in the outer region \citep{jimenez1999autonomous}. In
recent years, large-scale motions (LSMs), as well as very-large-scale motions (VLSMs) in the
logarithmic and outer regions, have been evidenced and received extensive attention
\citep{jimctr98,kim1999very,del2003spectra,del2004spectra,guala2006large,balakumar2007large,
hutchins2007evidence,monty2009comparison}. They are characterized by streamwise elongated large-scale
low- and high-speed regions and counter-rotating roll cells. The LSMs have a
streamwise length of approximately $h\sim3h$ \citep{adrian2007hairpin}, while the VLSMs
can be as long as $O(10h)$ and as wide as $O(h)$, where $h$ denotes
the outer length scale, such as the boundary layer thickness, half channel height, or pipe
radius. They not only carry more than half of the turbulent kinetic energy, but also a large
fraction of the Reynolds shear stress \citep{guala2006large}. They are active and
participate deeply in the dynamics of wall turbulence. However, the origin of the LSMs and VLSMs,
and their relation to near-wall structures are far from being settled, and further investigation
is needed.

A hypothesis proposed by \citet{kim1999very} attributes the generation of LSMs to the
alignment of hairpin vortices in the near-wall region. Vortex packets that carry a
low-momentum zone inside are assumed to induce much longer low-speed streaks as they
propagate downstream and lift the mean shear. Near-wall structures are suggested to provide
conditions for the LSM generation \citep{adrian2007hairpin}. This bottom-up hypothesis has
been evidenced by the subsequent researches in boundary layer flows \citep{lee2011very} and
pipe flows \citep{baltzer2013structural}.
\citet{Doohan21} also proposed a possible larger-scale eruption mechanism in the near-wall region, in terms of subharmonic streak instability.
This mechanism may lead to the formation of the wall-reaching part of high-speed large-scale streaks, although the result is limited to the buffer layer and the nearby scales.

Another hypothesis holds that the genesis of LSMs is the result of linear energy
amplification independent of the flow in the near-wall region. \citet{butfar93} and
\citet{del2006linear} revealed that the LSMs can be described well by the linear modes with
the largest transient growth. Pairs of large-scale counter-rotating roll cells, similar to
the widely documented statistical organization of turbulent flow, were observed in their
research, and similar findings by \citet{pujals2009note} and \citet{mckeon2010critical}
provide further support. Based on this hypothesis, a possible self-sustaining mechanism of
LSMs in the absence of near-wall structures was put forward, for example, by
\citet{flores2006effect}, who showed that the structures in the outer region remain
virtually unchanged after adding wall disturbances, regardless of the near-wall condition.
This agrees with the earlier conjecture by \citet{towns76} that the effect of wall roughness
on the structure of turbulence does not extend beyond a thin roughness layer.
\citet{hwang2010self} and \citet{hwang2017self} found that LSMs can self-sustain when the
small-scale motions are artificially quenched, agreeing with older conclusions from the
observation of large scales in large-eddy simulations of channel flow \citep{scov2001}.
Considering the similar self-sustaining process of near-wall structures, these results
suggest that dynamics of the structures at each relevant scale are mostly controlled by the
local mean shear through a linear amplification mechanism, rather than by the interaction
with larger or smaller scales. In fact \citet{miz:jim:13}, \citet{dong17} and \citet{Kwon21}
have shown that LSMs very similar to those in wall turbulence exist even in the absence of a
wall, and \citet{tue:jim:13} showed that small artificial changes in the off-wall shear in
channels have strong structural effects consistent with a local origins of the LSMs.

Beyond the different opinions on their origin, the influence of LSMs and VLSMs on the near-wall region
is widely documented. The footprint of large-scale structures reach deeply into the
near-wall region \citep{hoyas06,hutchins2007large}, and this top-down influence was
categorized into superposition and amplitude modulation by \citet{mathis2009large}.
Superposition is a linear process that refers to the contribution of the footprint of
large-scale motions to turbulent kinetic energy in the near-wall region
\citep{hoyas06,marusic2010high}. Amplitude modulation is a nonlinear process that describes
how small-scale turbulent fluctuations are enhanced in large-scale high-speed regions and
suppressed in low-speed ones. A predictive model for the streamwise velocity fluctuations in
the near-wall region has been proposed based on large-scale signals from the logarithmic layer
\citep{marusic2010predictive,mathis2011predictive}. Turbulent statistics are well predicted
by this model.
\citet{abe2004very} also proposed a possible top-down influence on the wall-shear stress fluctuations.
The positive and negative dominant regions of the streamwise wall-shear stress fluctuations were found to be corresponding with the high- and low-speed regions of the VLSMs, respectively, whereas the active regions of the spanwise wall-shear stress tend to concentrate under high-speed regions of the VLSMs, suggesting the possible top-down influence on the spanwise velocity fluctuations.

According to the hypothesis of hairpin packets, the near-wall region provides the
environment for the generation of LSMs. Combining the bottom-up hypothesis with the top-down
mechanism, it is conceivable that a co-supporting mechanism exists that involves the
near-wall structures and the LSMs of the outer flow. Such a hypothesis was
proposed by \citet{toh2005interaction}, who studied the correlation between spanwise motions
of near-wall streaks and outer large-scale structures in a streamwise-minimal channel. In it, the
large-scale circulations carry near-wall low-speed streaks toward up-washing regions, and
low-speed LSMs are generated by the merger and eruption of the near-wall streaks in the
areas where they concentrate. In down-washing regions, the outer-layer circulation
continually carries fluid towards the wall, enhancing the wall shear, and near-wall streaks
are created due to instabilities. The co-supporting cycle is completed when these two
processes are connected by the large-scale counter-rotating roll cells, and provides an
intuitive framework for the inner-outer interaction, in which the distribution of near-wall
streaks matches the LSMs in the spanwise direction.

However, the simulations of \citet{toh2005interaction} suffer from a low Reynolds number and a short computational domain.
There lacks solid mathematical description of the hypothesized co-supporting cycle, and whether the hypothesis holds is still an open question.
This motivates the present study, in which conditional statistics and numerical experiments are designed to further elucidate this issue.
Its main purpose is to examine and verify whether the co-supporting hypothesis survives in full-sized turbulent channel flows at higher Reynolds numbers
by checking its hypothesized bottom-up and top-down branches, as well as to quantify the impact of those branches on the flow.

The paper is organized as follows. The direct numerical simulation (DNS) data used in the
present study is described in \S\ref{sec:dns}. The trajectories of the spanwise locations of
the near-wall streaks in full-sized turbulent channels are analyzed in
\S\ref{sec:nearwall}. The top-down influence of the outer flow on the motion and density
of the streaks is examined and quantified in \S\ref{sec:drift} and \S\ref{sec:density}.
Numerical experiments are performed in \S\ref{sec:bottomup} to investigate the possible
bottom-up influence of the near-wall structures on the generation and maintenance of the outer
LSMs, and conclusions are offered in \S\ref{sec:conc}.

\section{DNS database of the turbulent channel flows}\label{sec:dns}

\begin{table}
  \begin{center}
\def~{\hphantom{0}}
  \begin{tabular}{lcccccccl}
       case   & $\Rey_\tau$ & $L_{x}$       & $L_{z}$     & $\Delta_x^+$  & $\Delta_z^+$  & $(\Delta_y)_{max}^+$   & $T{U}_{m}/h$   & Reference          \\[3pt]
       L550   & 547         & ${8\pi h}$    & ${4\pi h}$  & $13.4$        & $6.8$         & $6.7$                  & $172.4$        & \citet{del2003spectra}   \\
       M950   & 932         & ${2\pi h}$    & ${\pi h}$   & $11.5$        & $5.7$         & $7.7$                  & $276.8$        & \citet{lozano2014effect} \\
       M2000  & 2009        & ${2\pi h}$    & ${\pi h}$   & $12.3$        & $6.2$         & $8.9$                  & $113.9$        & \citet{lozano2014effect} \\
       M4200  & 4179        & ${2\pi h}$    & ${\pi h}$   & $12.8$        & $6.4$         & $10.7$                 & $90.9$         & \citet{lozano2014effect} \\[1ex]
       W535   & 535         & ${2\pi h}$    & ${4\pi h}$  & $11.8$        & $5.9$         & $8.8$                  & $200.0$        & \citet{deng2012influence}  \\
  \end{tabular}
\caption{Computational parameters, as detailed in the text. $\Delta_x$ and $\Delta_z$
are the resolution in the wall parallel directions, expressed in terms of Fourier modes, and
$\Delta_y$ is the collocation resolution in the wall-normal direction.
$T{U}_{m}/h$ is the sampling time period used in the statistics of \S\ref{sec:nearwall}.
The sampling time periods $T$ of the two numerical experiments in \S\ref{sec:bottomup} are $150h/{U}_{m}$ and $45h/{U}_{m}$, respectively.}
  \label{tab:cp}
  \end{center}
\end{table}

We use the DNS database of turbulent channel flow from \citet{del2003spectra} and
\citet{lozano2014effect}. The flow is established between two parallel plates separated by
$2h$, at Reynolds number, $\Rey_{\tau}=u_{\tau}h/\nu$, ranging from $547$ to $4179$. The
friction velocity $u_{\tau}$ and the kinematic viscosity $\nu$ define wall units, denoted by
a `+' superscript. Quantities without explicit units are assumed to be normalised with the
average bulk velocity, $U_m$, and with $h$. The streamwise, wall-normal, and spanwise
coordinates are $x$, $y$ and $z$, respectively, and the flow is assumed to be periodic in
the streamwise and spanwise directions, with periods $L_x$ and $L_z$, respectively. The
corresponding velocities are $u$, $v$ and $w$. The numerical code is Fourier-spectral in
these wall-parallel directions, and either uses Tchebychev polynomials or high-order compact
finite differences in the wall-normal one. The main numerical parameters are listed in table
\ref{tab:cp}, and the reader is directed to the original publications for further details.

The simulation in the last line in table \ref{tab:cp} will be used as reference in the series
of numerical experiments in \S\ref{sec:bottomup}, and will be discussed in more detail in
that section.

\section{Spanwise location and drift of the near-wall streaks}\label{sec:nearwall}

\begin{figure}
  \centering
  \begin{overpic}
  [scale=0.42]{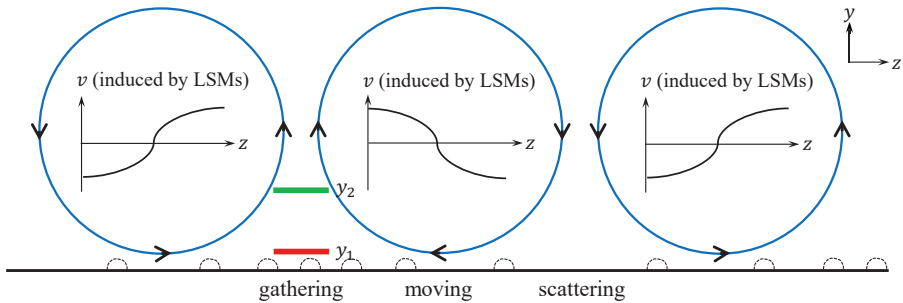}
  \end{overpic}
\caption{Sketch of the outer-inner interaction between large-scale streamwise rollers (blue circles)
and the near-wall low-speed streaks (dashed curves), according to the co-supporting
hypothesis in \citet{toh2005interaction}. The red and green lines denote the two
wall-parallel observation windows, $y=y_{1}$ and $y_{2}$,
respectively, used in \S\ref{sec:drift} and \S\ref{sec:density}.
}
\label{fig:sketch}
\end{figure}

As explained in \S\ref{sec:intro}, the \citet{toh2005interaction} co-supporting cycle
consists of bottom-up and top-down branches. See the sketch in figure \ref{fig:sketch}. The
current section is concerned with the top-down process, according to which the spanwise
drift of the near-wall streaks is driven by the large-scale streamwise rollers of the
outer flow. Streaks are driven away from regions of downwash, migrate sideways, and
concentrate in regions of upwash. Our first attempt at locating and defining streaks was to
use the method in \citet{toh2005interaction}. The spanwise location $z=\zeta(t,x_r,y)$ of a
meaningful low-speed streak is determined by
\begin{equation}
  u^{(2D)}(t,x,y,z)|_{x=x_{r},z=\zeta}<U(y),
  \label{eq:1}
\end{equation}
\begin{equation}
  \frac{\partial u^{(2D)}}{\partial z}(t,x,y,z)|_{x=x_{r},z=\zeta}=0,
  \quad
  \frac{\partial^{2}u^{(2D)}}{\partial z^{2}}(t,x,y,z)|_{x=x_{r},z=\zeta}>0,
  \label{eq:2}
\end{equation}
where the locally averaged instantaneous streamwise velocity,
\begin{equation}
  u^{(2D)}(t,x_{r},y,z)=\frac{1}{\Delta x}\int_{x_{r}-\Delta x/2}^{x_{r}+\Delta x/2}u(t,x,y,z)\,\mathrm{d}x,
  \label{eq:3}
\end{equation}
is used instead of $u$ itself.
The $x_{r}$ is the midpoint of the streamwise averaging interval and its determination will be discussed in details in the following.
A similar definition is used elsewhere for other velocity
components, $v^{(2D)}$ and $w^{(2D)}$. Condition (\ref{eq:1})
guarantees that $u^{(2D)}$ is slower than the mean velocity $U(y)$, and (\ref{eq:2}) ensures
that $\zeta$ is a spanwise minimum of $u^{(2D)}$.

The choice of the averaging interval $(x_r-\Delta x/2, x_r+\Delta x/2)$ is important.
\citet{toh2005interaction}, who analyzed a streamwise-minimal channel with $L_{x}^{+}=384$,
used the full streamwise period $\Delta x=L_x$ as their averaging length. This
length is shorter than most streaks, which were thus treated as being
infinitely long. Their evolution takes place in time rather than in space, and the same
group later showed that the length of the near-wall streaks in a full channel can be
substituted by their lifetime in a streamwise-minimal one \citep{toh2018MFU}.

We analyze in this section full-sized channels at substantially higher Reynolds numbers than in
either of the two papers above, and some modifications are needed to \r{eq:1}-\r{eq:3}. The
typical streamwise length of the near-wall streaks, $\ell_{xu}$, is  a few thousand wall
units \citep{hoyas06}, and averaging over longer segments risks mixing streaks and hiding
their spanwise meandering. We use in this paper $\Delta x^+\approx380\sim450$, which is short
enough to retain structural information, but long enough to differentiate the near-wall
streaks from velocity fluctuations of smaller scale. A posteriori tests show that the
results below are robust in the range $\Delta x^+ \approx 300\sim500$.

When $\Delta x=L_x$, the streamwise origin, $x_r$, of the observation box becomes
irrelevant, and the near-wall streaks in the streamwise-minimal flow in
\citet{toh2005interaction} form branching structures that tend to merge beneath large-scale
low-speed regions, supporting their assumed relation between LSMs and the spanwise drift of
the streaks. In larger channels, the position of the observation window is crucial.
\citet{kimh93} and \citet{del2009estimation} showed that the streamwise advection velocity
of most quantities in the buffer layer is $u_{ad}^+\approx 10$, and \citet{lozano2014time}
later showed that $u_{ad}^+\approx 8$ for the ejections associated with the low-speed
streaks. Farther from the wall, the small scales move approximately with the mean flow
velocity, with ejections also moving slightly more slowly than the flow \citep{lozano2014time}. For
example, ejections at $y^+=200$ move at $u_{ad}^+=16.7$ while $U(y)^+=18.2$. Centering on
the buffer layer, we shall see below that the spanwise drift velocity of the streaks is of
the order of $u_\tau$, so that, if the observation window were fixed to the wall, streaks
would only be observed over times of the order of $\Delta t^+\approx \ell_{xu}^+/8\approx
100$, during which they would drift at most by $\delta z^+\approx 100$. This is too little
to characterize the interactions in which we are interested, which involve VLSMs whose width
is $\Delta z\approx h$. Therefore, we track individual streaks by moving the
window with their average advection velocity, $x_r=x_{r0} +u_{ad} t$, where $x_{r0}$ is the
initial window position.

\begin{figure}
    \centering
    \subfigure{
    \begin{overpic}
    [scale=0.27]{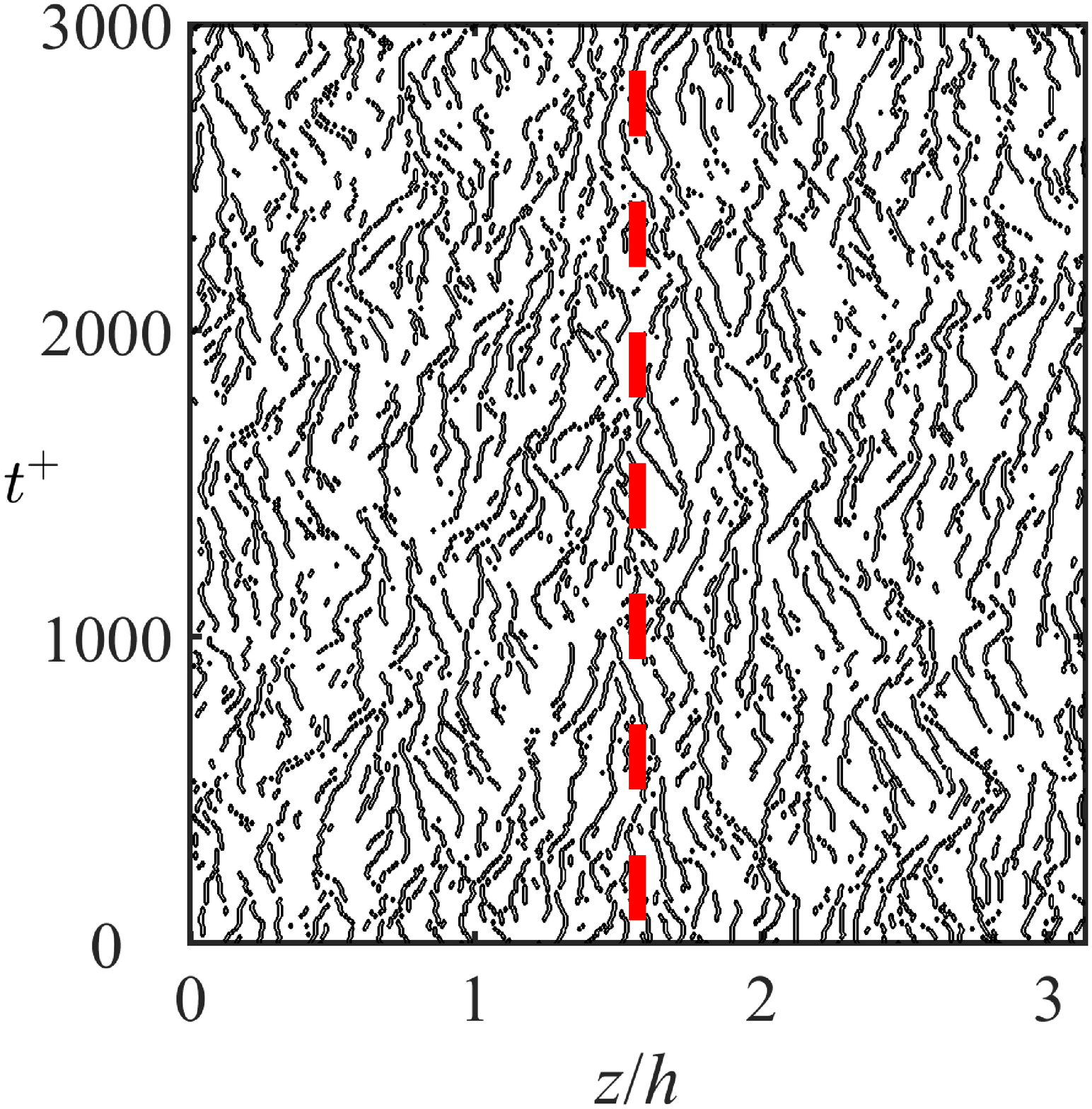}%
    \put(-5,94){(\textit{a})}%
    \end{overpic}
\hspace*{3mm}
    \begin{overpic}
    [scale=0.27]{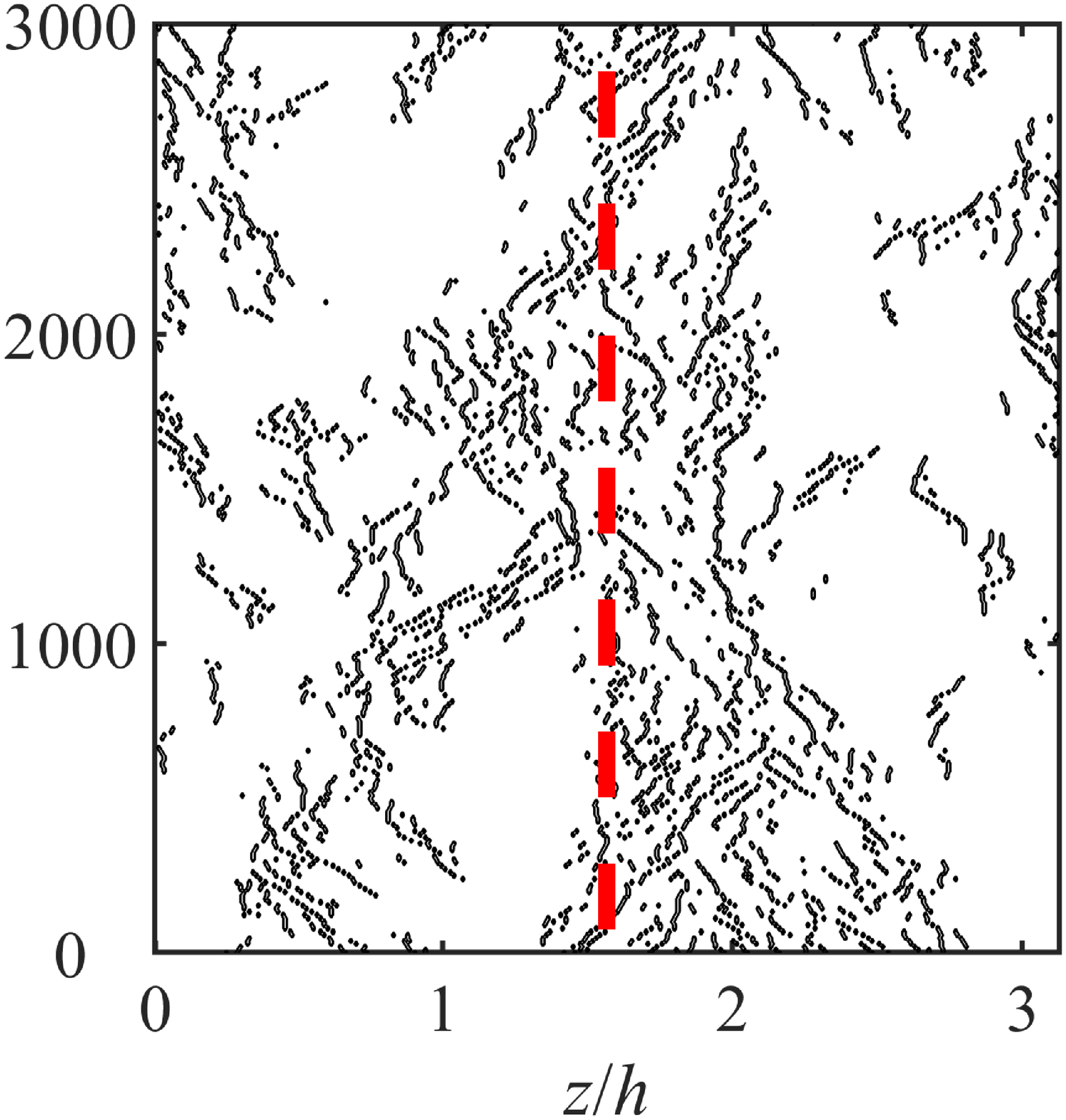}
    \put(-5,94){(\textit{b})}
    \end{overpic}
    }
\caption{Time history of the spanwise locations of low-speed streaks in a full-sized channel
(M950). Locations of the low-speed streaks are determined by the condition \r{eq:1}-\r{eq:3}.
(\textit{a}) $y^{+}=5$ and (\textit{b}) $y^{+}=200$. The red dashed line denotes the
averaged location of a large-scale streak.}
    \label{fig:zitaM950}
\end{figure}

\begin{figure}
    \centering
    \subfigure{
    \begin{overpic}
    [scale=0.27]{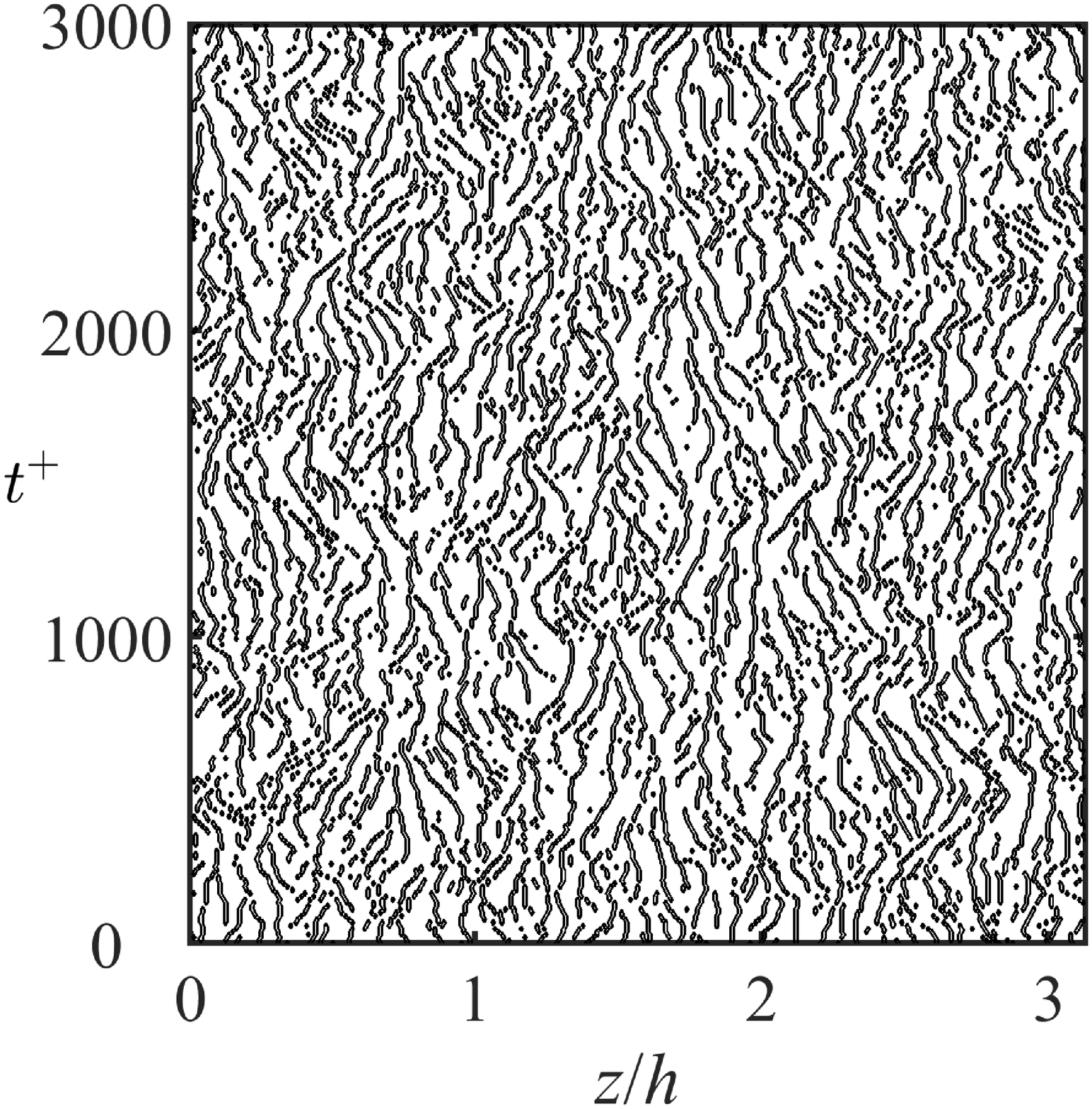}%
    \put(-5,94){(\textit{a})}%
    \end{overpic}
\hspace*{3mm}
    \begin{overpic}
    [scale=0.27]{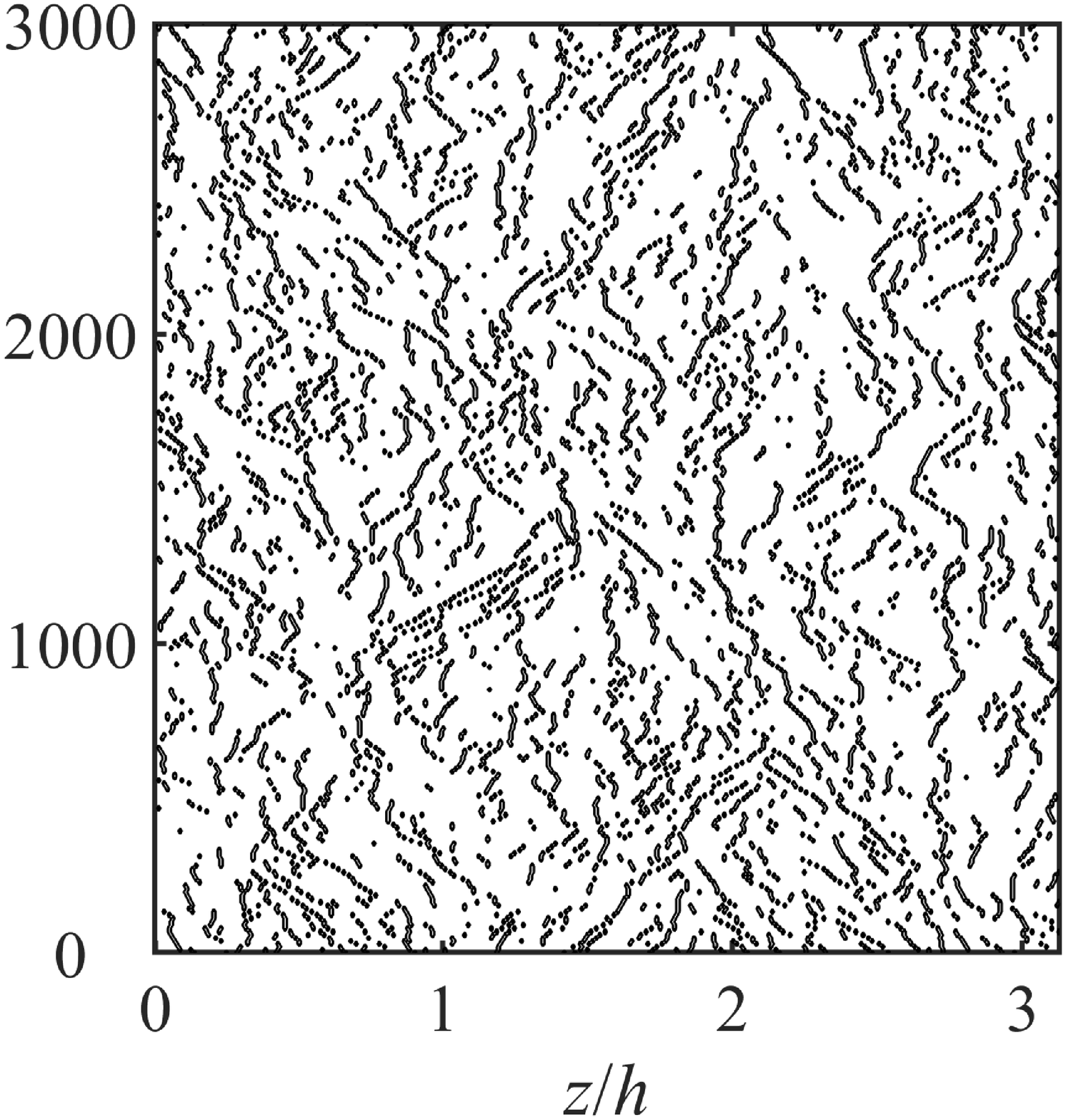}
    \put(-5,94){(\textit{b})}
    \end{overpic}
    }
\caption{Time history of the spanwise locations of low-speed streaks in a full-sized channel
(M950) without the criterion \r{eq:1}. (\textit{a}) $y^{+}=5$ and (\textit{b}) $y^{+}=200$.}
\label{fig:zitaM950new}
\end{figure}

A typical time history of the spanwise location of low-speed streaks at $y^{+}=5$ is
displayed in figure \ref{fig:zitaM950}(\aaa), using $u_{ad}^+=8$ as representative of the
near-wall region. Each line in the figure represents the trajectory of a single low-speed
streak, and it is clear from the figure that they substantially drift spanwise. Their mean
spanwise spacing, about $100$ wall units, agrees well with the known streak spacing near the
wall \citep{smet83}. \citet{toh2005interaction} classify the trajectories that they observe
in streamwise-minimal channels into dominant branches with long lifetimes (of the order of
1000--3000 wall units), and subordinate ones with shorter lifetimes that tend to merge into
the dominant ones. The lifetime of such a fragile object as a near-wall streak is hard to
define, but Lagrangian scales of 300--1000 temporal wall units were obtained in the buffer
layer by \citet{jim05}, \cite{flores2010hierarchy} and \cite{toh2018MFU}. The lifetime of the branches in
figure \ref{fig:zitaM950}(\aaa) is of the same order, 200--800 wall units, and few of them
last beyond 1000 viscous time units. They  thus correspond to the subordinate branches
in  \citet{toh2005interaction}.

Figure \ref{fig:zitaM950}(\bbb) is drawn at $y^{+}=200$, using $u_{ad}^+=16.7$, and
represents the logarithmic region. The trajectories cluster in a wide band around $z/h=1.7$,
marked by a red dashed line, which appears to be one of perhaps two large streaks that
dominate the flow at that level. The interpretation of this figure needs some comment.
Although it is known that outer-layer low-speed regions in channels contain more small-scale vortices
than high-speed ones \citep{tana04}, which could be interpreted as
supporting the gathering action hypothesized by \citet{toh2005interaction}, the clustering
in figure \ref{fig:zitaM950}(\bbb) is probably an artefact of the detection criterion
\r{eq:1}. Low-velocity streaks are only recognized as such when they ride on a deeper large
streak that lowers their velocity below the mean profile, and what we see in the figure is
probably the root of an attached streak centred farther from the wall.

The influence of criterion \r{eq:1} should thus be further discussed. Figure
\ref{fig:zitaM950new} repeats figure \ref{fig:zitaM950} without using \r{eq:1}. At
$y^{+}=5$, condition \r{eq:1} discards part of the scattered dots, which reappear in figure
\ref{fig:zitaM950new}, but there are also more coherent streaks in the new figure than in
the old one. The spanwise drift of the streaks is present in both figures, but it is evident
from comparing them that the streak density measured by \r{eq:1}--\r{eq:2} may be influenced
by \r{eq:1}. For the result at $y^{+}=200$ in figures \ref{fig:zitaM950}(b) and
\ref{fig:zitaM950new}(b), the condition \r{eq:1} takes a dominant role, and rejects about
half the local minima. When it is removed, the gathering action suggested by the
co-supporting is still observed, although less clearly than before, but it is not reflected in
the streak density. Figure \ref{fig:zitaM950new} shows an almost uniform distribution of
small streaks, instead of only the part riding on the larger outer streak. Only condition
\r{eq:2} will be used to define the near-wall streaks in the quantitative analysis in the
next two sections.

\subsection{Spanwise drift of the near-wall streaks, versus the outer eddies}\label{sec:drift}

Although figures \ref{fig:zitaM950} and \ref{fig:zitaM950new} clearly indicate that the
streaks move spanwise, relating their drift to the outer large-scale circulations requires
quantitative analysis. The premise of the top-down branch of the co-supporting cycle in
\citet{toh2005interaction} is that the near-wall streaks are scattered away from the
down-washing regions of the large-scale circulations, and gather in the up-washing ones, as
sketched in figure \ref{fig:sketch}. This implies a positive correlation between the
spanwise derivative, $\partial v/\partial z$, of the wall-normal velocity of the outer flow
and the spanwise drift of the near-wall streaks. A positive spanwise streak velocity near
the wall would correspond to a counter-clockwise large-scale circulation, in which
$\partial v/\partial z>0$ near the $z$-positions of the roller centers, as depicted by the first and
third rolls in figure \ref{fig:sketch}. A negative streak drift velocity would be associated with
$\partial v/\partial z<0$.

To quantify how streaks move as a function of their location, $(t,x_r,y_1,z_s)$, we track
the evolution of patterns of $u^{(2D)}$, using a method similar to particle image
velocimetry (PIV). Consider a one-dimensional interrogation window,
$z_{s}-\Delta\zeta/2<z<z_{s}+\Delta\zeta/2$, which is advected streamwise with
the velocity discussed in the previous section. The streak displacement after $\Delta t$ is
defined as the position, $\delta z_{max}$, of the maximum of the correlation
\begin{equation}
R_{u^{(2D)}}(\delta z,\Delta t)=\frac{1}{(I_{0}I_{1})^{\frac{1}{2}}}\int_{z_{s}-\Delta\zeta/2}^{z_{s}+\Delta\zeta/2}u^{(2D)}(t,x_r,y_1,z)u^{(2D)}(t+\Delta t,x_r+u_{ad}\Delta t,y_1,z+\delta z)\,\mathrm{d}z,
\label{eq:4}
\end{equation}
where $I_{0}$ and $I_{1}$ are, respectively, the mean squares of $u^{(2D)}(t,x_r,y_1,z)$ and
$u^{(2D)}(t+\Delta t,x_r+u_{ad}\,\Delta t,y_1,z+\delta z)$, averaged over the same window as
in \r{eq:4}. In addition, we only accept maxima for which $R_{u^{(2D)}}(\delta
z_{max})>0.8$. The spanwise advection velocity is defined as $w_{s}=\delta z_{max}/\Delta
t$, and is assigned to the window centre $(x_r, z_s)$. The following discussion focuses on
the near-wall streaks at $y_1^+=13$. It uses $\Delta\zeta^{+}\approx 50$, and an advection
velocity of the interrogation window $u_{ad}^+=8$, as in figure \ref{fig:zitaM950}(\aaa).
The results are robust in the range $\Delta\zeta^{+}\in \left[30,70\right]$, and the effect
of $\Delta t^+$ and $u_{ad}^+$ is discussed in appendix \ref{sec:PIV}.

\begin{figure}
  \centering
  \subfigure{
  \begin{overpic}
    [scale=0.29]{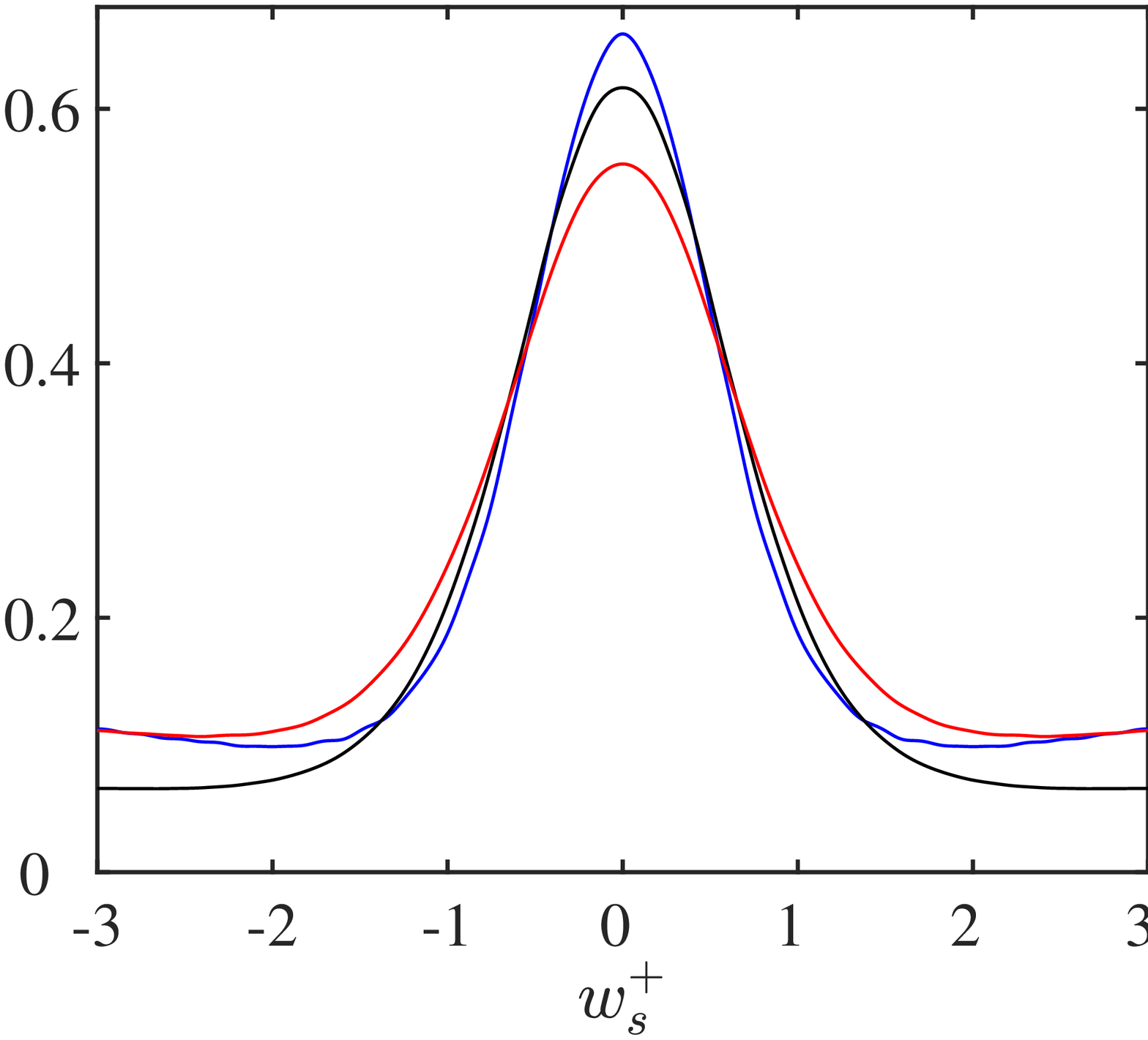}
    \put(-1,77){(\textit{a})}
  \end{overpic}
  \begin{overpic}
    [scale=0.29]{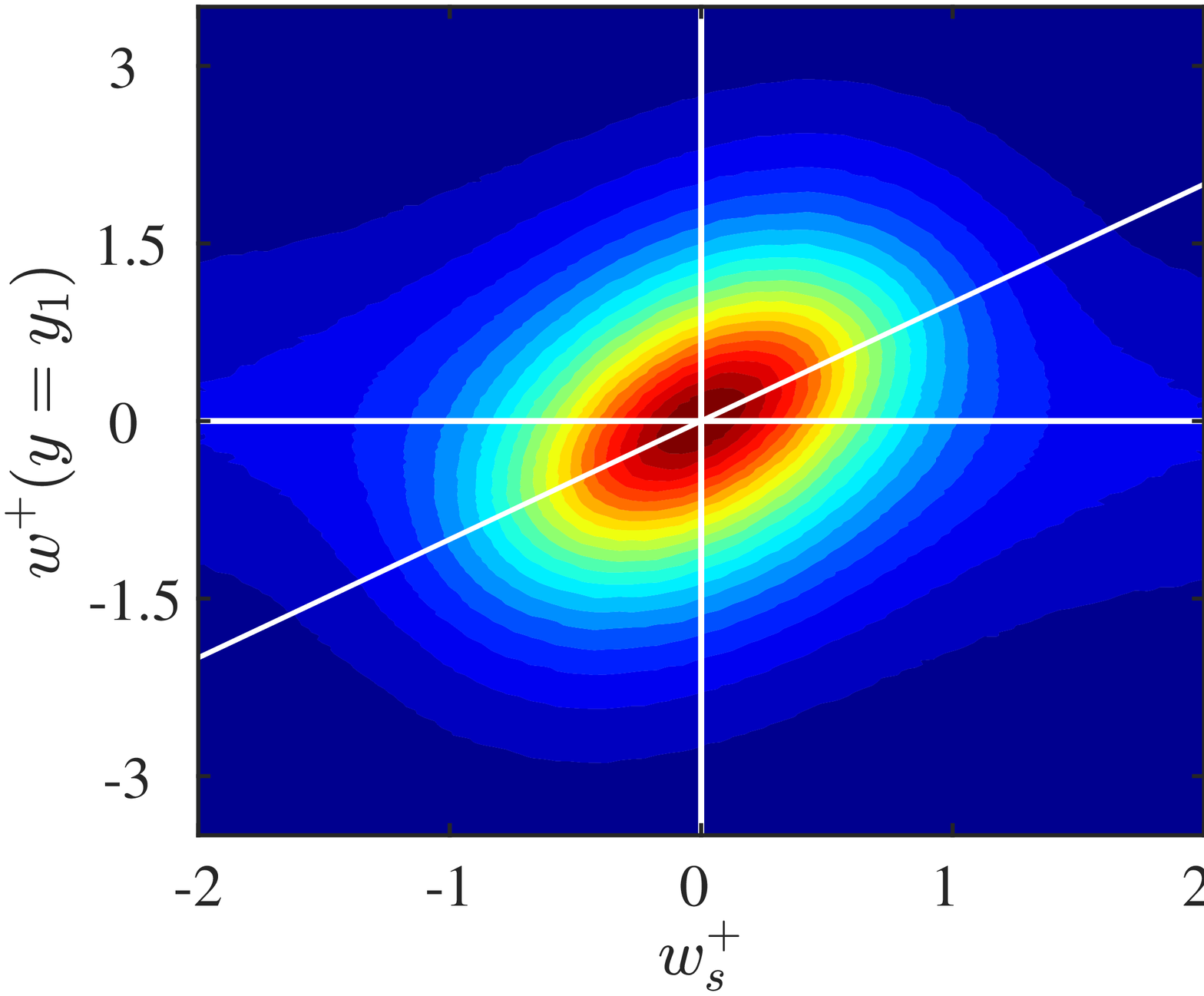}
    \put(-1,77){(\textit{b})}
  \end{overpic}
  }
  \DeclareRobustCommand\mylabela{\tikz[baseline]{\draw[solid, blue, thick] (0,0.5ex) -- (0.8,0.5ex);}}
  \DeclareRobustCommand\mylabelb{\tikz[baseline]{\draw[solid, black, thick] (0,0.5ex) -- (0.8,0.5ex);}}
  \DeclareRobustCommand\mylabelc{\tikz[baseline]{\draw[solid, red, thick] (0,0.5ex) -- (0.8,0.5ex);}}
\caption{%
 (\textit{a}) Probability density function of $w_{s}^{+}$ at different
$\Rey_{\tau}$ with $\Delta t^{+}\approx 20$, and $u_{ad}^+=8$. \mylabela, Case W535;
\mylabelb, M950; \mylabelc, M2000.
(\textit{b}) Joint probability density function of the streak spanwise advection velocity,
$w_{s}^{+}$, and the spanwise velocity fluctuations, $w^+$ at $y_{1}^{+}=13$. Case M950, and
$\Delta t^{+}=20$. The white diagonal is $w^{+}=w_{s}^{+}$.}
\label{fig:pdfws}
\end{figure}

The probability density function (PDF) of the drift velocity at different $\Rey_{\tau}$
shown in figure \ref{fig:pdfws}(\textit{a}) is concentrated approximately in the range
$w_s^+=[-2,2]$, independently of the Reynolds number. This is the same order of magnitude as
the spanwise velocity fluctuations of the flow, and the two quantities are closely related.
The joint PDF of $w_{s}^{+}$ and $w^{+}$ at $y_{1}^{+}=13$ is displayed in figure
\ref{fig:pdfws}(\textit{b}) for case M950. It shows a clear correlation between the two
variables, providing direct evidence that the drift of the streaks is not due to their
internal dynamics, but to the ambient flow.

\begin{figure}
  \centering
  \subfigure{%
  \begin{overpic}
    [scale=0.29]{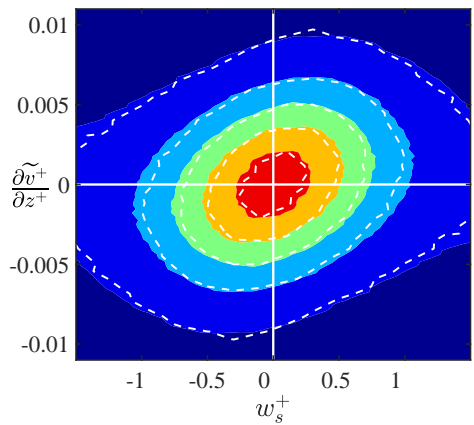}
  \end{overpic}
  }

  \DeclareRobustCommand\mylabela{\tikz[baseline]{\draw[solid, blue, thick] (0,0.5ex) -- (0.8,0.5ex);}}
  \DeclareRobustCommand\mylabelb{\tikz[baseline]{\draw[solid, black, thick] (0,0.5ex) -- (0.8,0.5ex);}}
  \DeclareRobustCommand\mylabelc{\tikz[baseline]{\draw[solid, red, thick] (0,0.5ex) -- (0.8,0.5ex);}}
\caption{%
Joint probability density function of the spanwise drift velocity $w_{s}^{+}$ at $y_1^+=13$
and $\partial\widetilde{v}^{+}/\partial z^{+}$ at $y_2^+=200$. Shaded: $w_{s}^{+}$ by PIV; White
dashed lines: $w_{s}^{+}$ at the streak centers. Contour levels are $0.1 (0.2) 0.9$
of the maximum probability density. Case M950, $\Delta t^+\approx 20$, and $\Delta z^{+}= 214$.}
\label{fig:pdft5}
\end{figure}

We next turn our attention to the correlation of the drift velocity with the structures
of the outer flow, represented by the wall-normal velocity at
$y_2>y_1$, smoothed with a top-hat filter to isolate the large-scale component,
\begin{equation}
\widetilde{v}(x_0,y_2,z_0)=\frac{1}{\Delta z}\,
\intop_{z_{0}-\frac{1}{2}\Delta z}^{z_{0}+\frac{1}{2}\Delta z} v^{(2D)}(x_0,y_{2},z)\,\mathrm{d}z,
\label{eq:5}
\end{equation}
where $v^{(2D)}$ is defined as in \r{eq:3}. A typical joint PDF is shown in figure
\ref{fig:pdft5}. The abscissae are $w_s$ at $y_1^+=13$, and the ordinates are the spanwise
derivative of $\widetilde{v}$ at $y_2^+=200$. It is evident that the PDF is preferentially
aligned to the first and third quadrants, i.e., that a positive spanwise drift of the
streaks tends to occur beneath positive $\partial\widetilde{v}/\partial z$, and vice versa,
in agreement with the top-down model. To verify that the continuous velocity field obtained
by the PIV method truly represent the drift of the streaks, the joint PDF, using $w_s^+$
collected at the points identified by (\ref{eq:2}) as the streak centers, are also shown by
the white contours in figure \ref{fig:pdft5} for comparison. The differences with the PIV
results are negligible.

The mutual dependence of the two quantities can be quantified by the correlation coefficient
\begin{equation}
R(w_s,\frac{\partial\widetilde{v}}{\partial z})=\frac{\sum w_{s}\frac{\partial\widetilde{v}}{\partial z}}%
{\left[\sum w_s^2\sum (\frac{\partial\widetilde{v}}{\partial z})^{2}\right]^{\frac{1}{2}}}.
\label{eq:6}
\end{equation}
Appendix \ref{sec:PIV} shows that $R(w_s,\frac{\partial\widetilde{v}}{\partial z})$ depends
on the PIV parameters, and on the wall-normal positions $y_2$ of the outer-flow. For the case in
figure \ref{fig:pdft5}, the correlation of the two quantities is
$R_{w_s,\partial\widetilde{v}/\partial z}\approx 0.2$, but it can be raised to the order of $0.4$ as
$y_2$ changes. See Appendix A for the details.

In fact, one of the functions of the window used to define $w_s$ and $\widetilde{v}$ is to
highlight how the correlation depends on the spanwise wavelength and on the distance from the wall.
The top-down correlations considered by \citet{toh2005interaction} and
\citet{toh2018MFU} were restricted to outer scales with $\lambda_{z}\sim h$, but it makes
sense to also consider interactions with intermediate scales presenting in the logarithmic
layer, for which $100\lesssim \lambda_{z}^{+}$ and $\lambda_{z}\lesssim h$. Consider the
inertia tensor of the joint PDF of $w_s$ and the unfiltered $\partial v/\partial z$, defined
as
\begin{equation}
\mathbf{I}\left(w_s, \tfrac{\partial v}{\partial z}\right) =
\left(\begin{array}{cc}
I_{w_{s}w_{s}} & I_{w_{s}\frac{\partial v}{\partial z}}\\
I_{\frac{\partial v}{\partial z}w_{s}} & I_{\frac{\partial v}{\partial z}\frac{\partial v}{\partial z}}
\end{array}\right),
\label{eq:inert1}
\end{equation}
where
\begin{equation}
I_{ab} =\frac{1}{L_x L_z}\, \iint a(x,z) b(x,z) \,\mathrm{d}z \,\mathrm{d}x.
\label{eq:inert2}
\end{equation}

To isolate the spanwise scale, express each variable as its Fourier transform along $z$,
using $\widehat{\varphi}(t,x,y,k_{z})$ to represent the Fourier coefficient of
$\varphi(t,x,y,z)$ at the spanwise wavenumber $k_{z}$. The different moments can be
expressed as integrals over $k_{z}$,
\begin{equation}
I_{w_{s}w_{s}}=\intop\widehat{w_{s}}\widehat{w_{s}}^{*}\,\mathrm{d}k_{z},
\end{equation}
\begin{equation}
I_{\frac{\partial v}{\partial z}\frac{\partial v}{\partial z}}=\intop ik_{z}\widehat{v}(ik_{z}\widehat{v})^{*}\mathrm{d}k_{z}=\intop k_{z}^{2}(\widehat{v}\widehat{v}^{*})\mathrm{d}k_{z},
\end{equation}
\begin{equation}
I_{w_{s}\frac{\partial v}{\partial z}}=I_{\frac{\partial v}{\partial z}w_{s}}=\intop\mathrm{Re}(ik_{z}\widehat{v}\widehat{w_{s}}^{*})\mathrm{d}k_{z}=\intop\mathrm{Re}(\widehat{w_{s}}(ik_{z}\widehat{v})^{*})\mathrm{d}k_{z}=\intop-k_{z}\mathrm{Im}(\widehat{v}\widehat{w_{s}}^{*})\mathrm{d}k_{z},
\end{equation}
where the `$*$' superscript denotes complex conjugation, and averaging over $x$ is implied everywhere.
A spectral inertia tensor can then be defined for each $k_z$ as
\begin{equation}
\mathbf{I}\left(w_s, \tfrac{\partial v}{\partial z}; k_z\right) =\left(\begin{array}{cc}
\widehat{w_{s}}\widehat{w_{s}}^{*} & -k_{z}\mathrm{Im}(\widehat{v}\widehat{w_{s}}^{*})\\
-k_{z}\mathrm{Im}(\widehat{v}\widehat{w_{s}}^{*}) & k_{z}^{2}(\widehat{v}\widehat{v}^{*})
\end{array}\right).
\label{eq:8}
\end{equation}
We are interested in the relation of $k_z$ with the height at which $\widehat{v}(k_{z})$ has the strongest influence on $w_s$.
The strength of this influence can be quantified by the inclination angle $\theta$ of the principal axes of the joint PDF inertia ellipse, and $\theta=\arctan|a_2/a_1|$ can be calculated from the leading eigenvector $(a_{1},a_{2})$ of the inertia tensor.
Here we can define a rescaled tensor that avoids derivatives by removing the $k_z$ factors from $\mathbf{I}$, for $k_z$ in \r{eq:8} not only bring dimensional complications, but also has a scale-dependent influence on the principal axes.
\begin{equation}
\mathbf{I}\left(w_s, v; k_z\right) =\left(\begin{array}{cc}
\widehat{w_{s}}\widehat{w_{s}}^{*} & -\mathrm{Im}(\widehat{v}\widehat{w_{s}}^{*})\\
-\mathrm{Im}(\widehat{v}\widehat{w_{s}}^{*}) & \widehat{v}\widehat{v}^{*}
\end{array}\right).
\label{eq:9}
\end{equation}
This tensor directly relates the velocities at the two wall distances at scale $k_z$, and
has the advantage of balancing their magnitudes and dimensions. Note that the imaginary part
in the off-diagonal elements of \r{eq:9} has the effect of translating spanwise one of the variables
by a quarter wavelength with respect to the other, so that $\mathbf{I}(w_s, v; k_z)$ describes
the correlation of the streak drift velocity with the shifted wall-normal velocity
in the outer flow.

\begin{figure}
  \centering
  \subfigure{
  \begin{overpic}
    [scale=0.26]{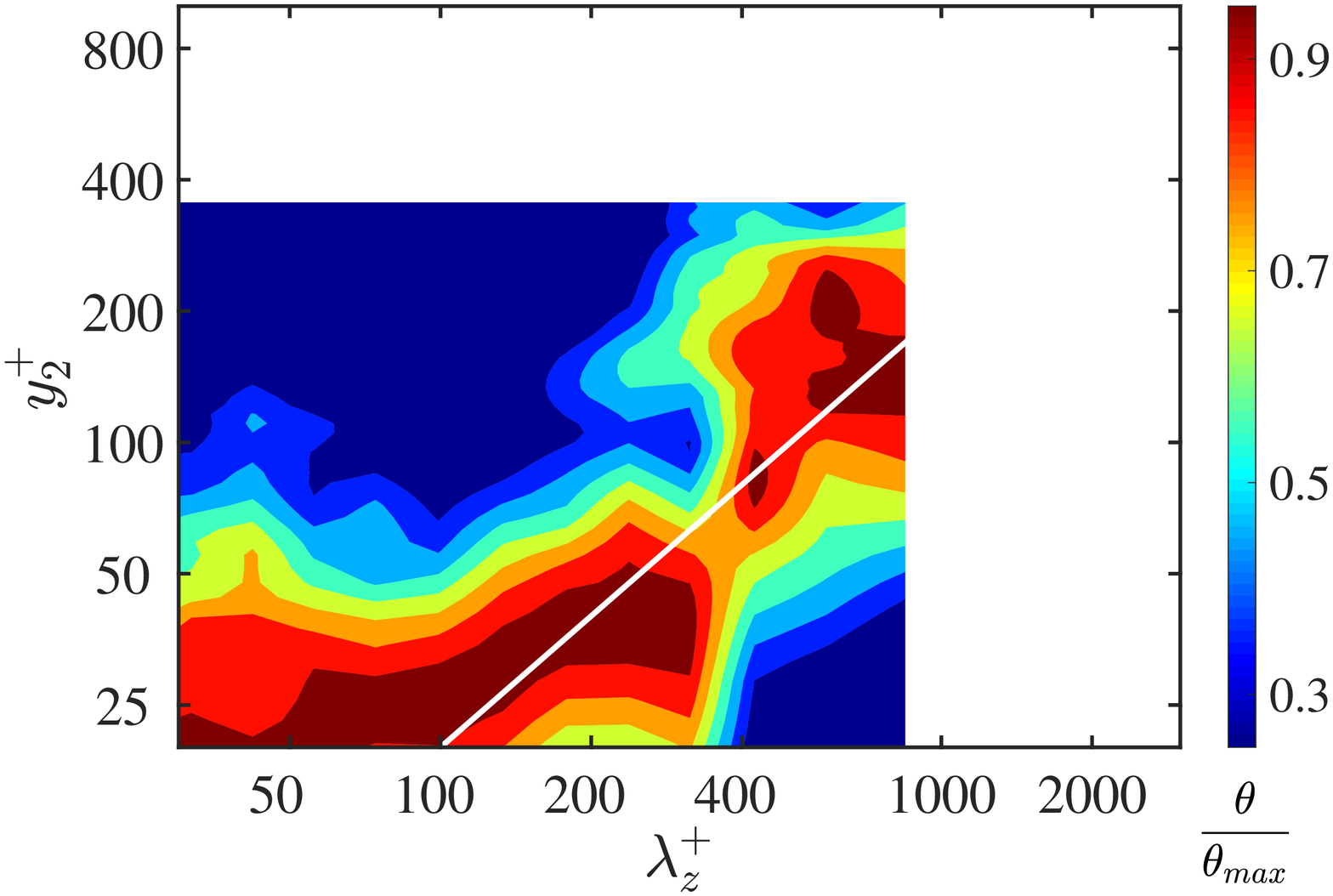}
    \put(-1,70){(\textit{a})}
  \end{overpic}
  \begin{overpic}
    [scale=0.26]{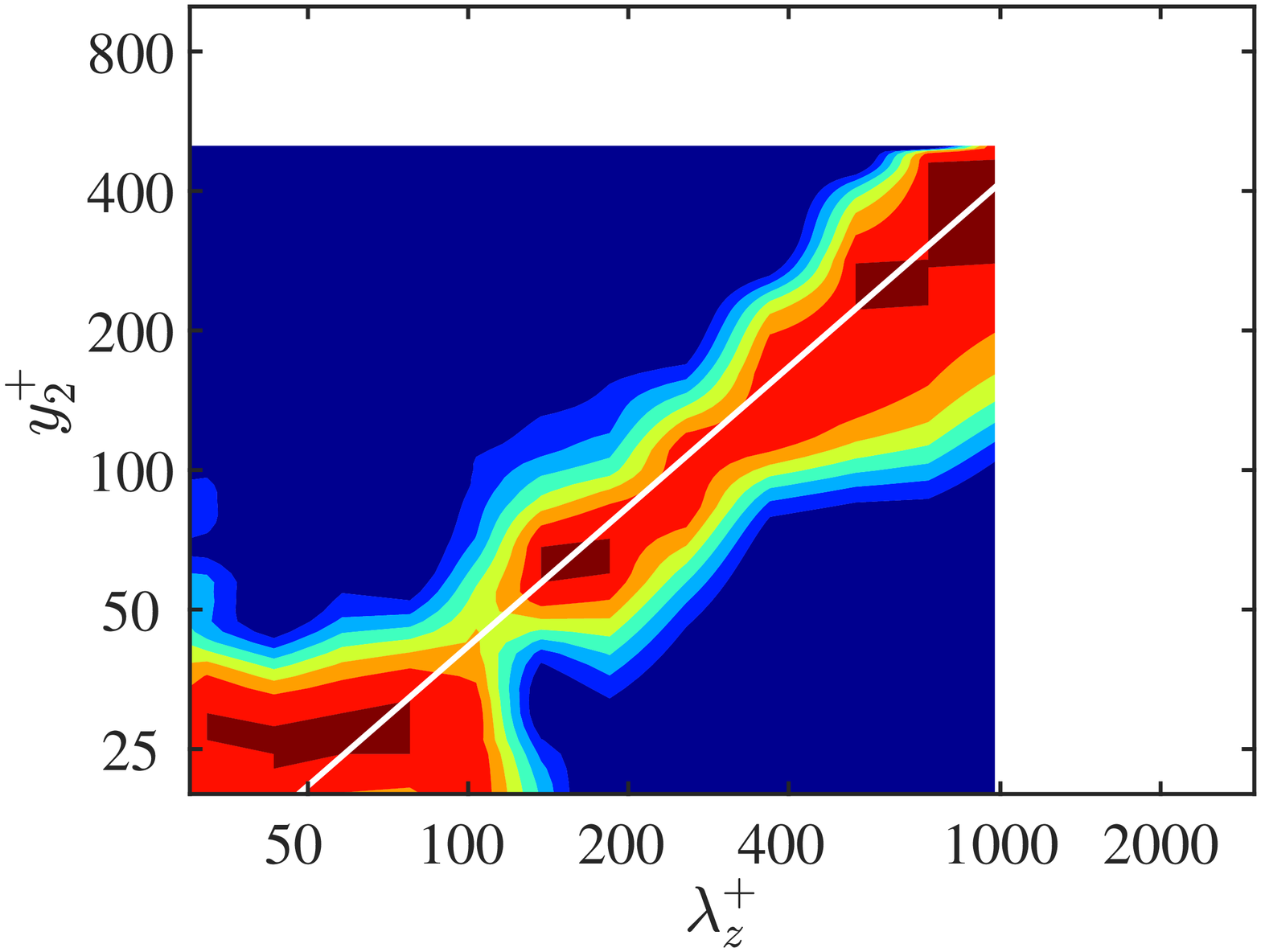}
    \put(-1,70){(\textit{b})}
  \end{overpic}
  }
  \subfigure{
  \begin{overpic}
    [scale=0.26]{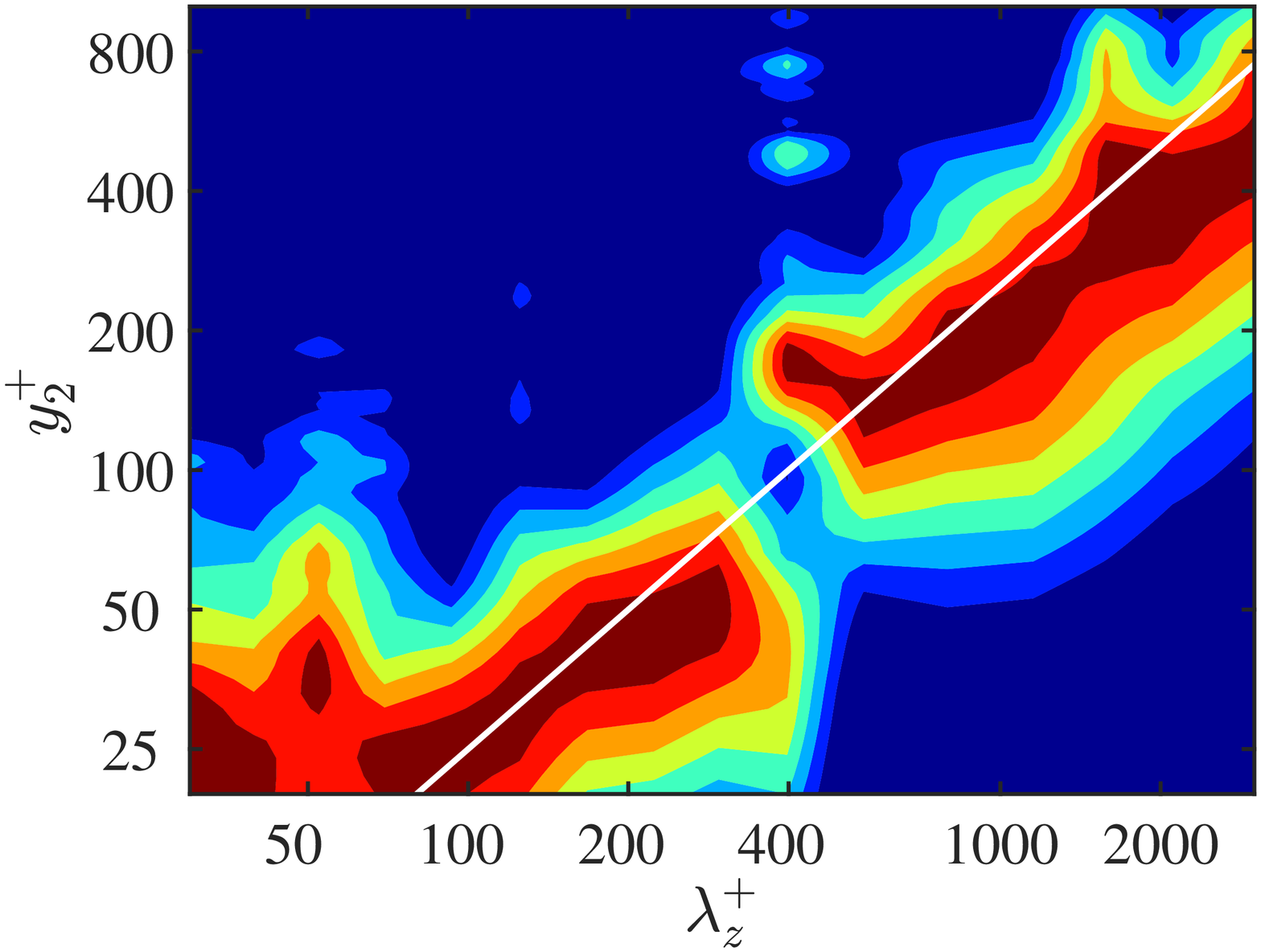}
    \put(-1,70){(\textit{c})}
  \end{overpic}
  }

  \caption{
Inclination angle, $\theta/{\theta}_{max}$, of the leading eigenvector of the correlation
tensor \r{eq:9}, as function of $y_{2}$ and of the spanwise wavelength $\lambda_{z}$, where
${\theta}_{max}$ is the maximum of $\theta$ at the corresponding wavelength. (\textit{a}) W535.
(\textit{b}) M950. (\textit{c}) M2000.
}
\label{fig:kiRet}
\end{figure}

The inclination angle $\theta$ determined by the second and first components of the leading eigenvector of \r{eq:9},
with $0<\theta<\pi/2$, quantifies the magnitude of the $v$ fluctuations at
$y_2$, relative to the drift velocity of the near-wall streaks with which they correlate.
The location, $y_2$ of the maximum for each wavelength, $\lambda_z=2\pi/k_z$, can be taken to
represent the distance from the wall of the centre of the rollers with that spanwise
dimension. It is displayed in figure \ref{fig:kiRet}, which clearly shows that rollers are
most effective at advecting streaks over distances similar to their own height (of the order
of $\lambda_z\approx 4y_2$). For example, the maximum correlation in figure
\ref{fig:RM950tutau} in appendix \ref{sec:PIV} is obtained when the wall-normal velocity is
smoothed with a filter for which $\Delta z^+\approx 200$ at $y_2^+=100$. Since the effect of
the eddies responsible for this correlation reaches, by definition, the near-wall region,
the self-similar dependence of their size is one more demonstration of the attached-eddy
model of \citet{town61}, and proves that the buffer-layer streaks are not controlled by
eddies of a single size, but by the whole hierarchy of attached structures of the
logarithmic layer.

\subsection{Near-wall streak density and the outer region}\label{sec:density}

Having established the influence of the outer structures on the spanwise drift of the streaks
of the buffer-layer, it remains to be shown whether this drift results in the
modulation of the density of streaks. This assumption underlies the second, bottom-up,
branch of the co-supporting hypothesis of \citet{toh2005interaction}, which proposes that
the accumulation of low-speed streaks below an existing outer-flow ejection leads to some
kind of collective instability that reinforces the ejection. At first sight, this conclusion
is reasonable, because the streaks converge where $\widetilde{v}(y_2)>0$ and diverge
where $\widetilde{v}(y_2)<0$. However, the density of streaks not only depends on their motion,
but also on the rate at which they are born and disappear, and it should be clear from the
discussion in \S\ref{sec:nearwall} that the longest reasonable lifetime of a streak is
of the same order as the time required for it to drift across distances of order $h$. In fact, a
cursory inspection of figure \ref{fig:zitaM950new} shows that most streaks do not live long
enough to cross spanwise distances of the order of the width of the largest VLSMs, but,
since we saw at the end of \S\ref{sec:drift} that streaks also interact with narrower
structures closer to the wall, we need to determine whether the same conclusion applies to
these structures. In this section we examine directly the correlation of the outer
$\widetilde{v}$ with the density of buffer-layer streaks.

To quantify this correlation, we again define two spanwise windows, one at $y_1^+=13$ in the
buffer layer, and another at $y_2$ in the outer region. Both windows have the same width,
$\Delta z$, and are centred at the same wall-parallel location $(x_0, z_0)$, but they serve
different purposes. The inner window is used to compute the streak density by counting
minima of $u^{(2D)}$, as in \r{eq:2}. The outer window is used to compute the smoothed
$\widetilde{v}$, as in \r{eq:5}. The streamwise interval used to compute $u^{(2D)}$ and
$v^{(2D)}$ is the same as in \S\ref{sec:nearwall}, $\Delta x^+=400$, but several $\Delta z$
are used to test their effect on the results. A minimum value for $\Delta z$ can be
estimated from physical arguments. Since the spacing of the buffer-layer streaks is known to
be about 100 wall units \citep{kline1967structure,smet83}, and the hypothesis in
\citet{toh2005interaction} is that at least two streaks need to merge to generate a larger-scale
`eruption', it is reasonable to choose $\Delta z^+\gtrsim 200$. Statistics are compiled by
scanning $x_0$, $z_0$ and time. The number of samples in each direction is summarised in
table \ref{tab:n_sample}. The total number of samples used for each Reynolds number always
exceeds $4\times10^{6}$.

\begin{table}
  \begin{center}
\def~{\hphantom{0}}
  \begin{tabular}{lcccc}
                   & L550      & M950      & M2000     & M4200     \\[3pt]
       $n_{x}$     & 30        & 15        & 30        & 61        \\
       $n_{z}$     & 1470      & 710       & 1470      & 3010      \\
       $n_{t}$     & 110       & 575       & 430       & 40        \\
       $N$         & $4.85\times10^{6}$ & $6.12\times10^{6}$ & $1.90\times10^{7}$ & $7.22\times10^{6}$\\
  \end{tabular}
\caption{Number of samples used for the streak statistics. $n_{x}$, $n_{z}$, and $n_{t}$
denote sample numbers in the $x$, $z$, and $t$ directions, and $N=n_{x} n_{z} n_{t}$
is the total number of samples.}
  \label{tab:n_sample}
  \end{center}
\end{table}


\begin{figure}
  \centering
  \subfigure{
  \begin{overpic}
    [scale=0.3]{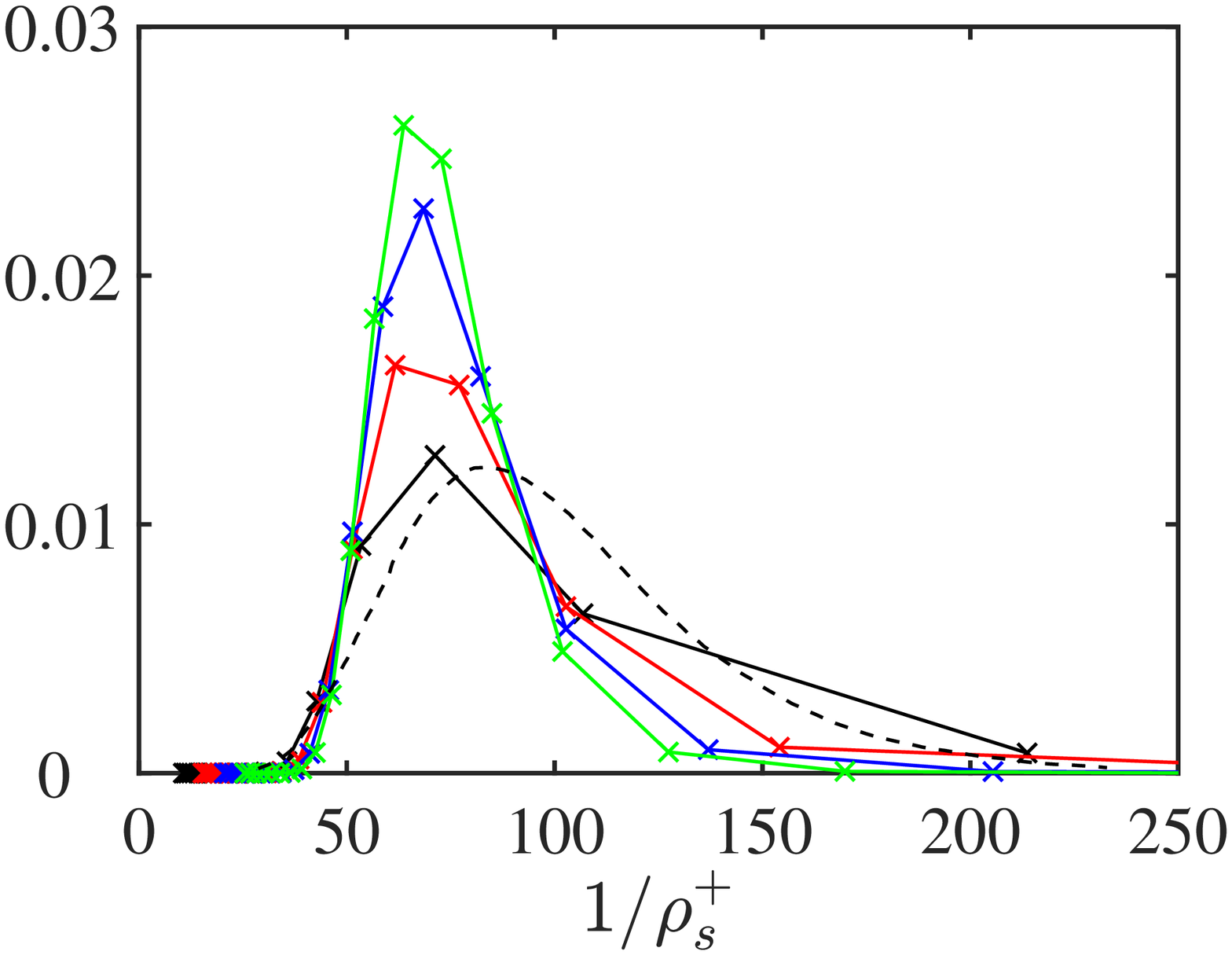}
    \put(-6,70){(\textit{a})}
  \end{overpic}
  \begin{overpic}
    [scale=0.3]{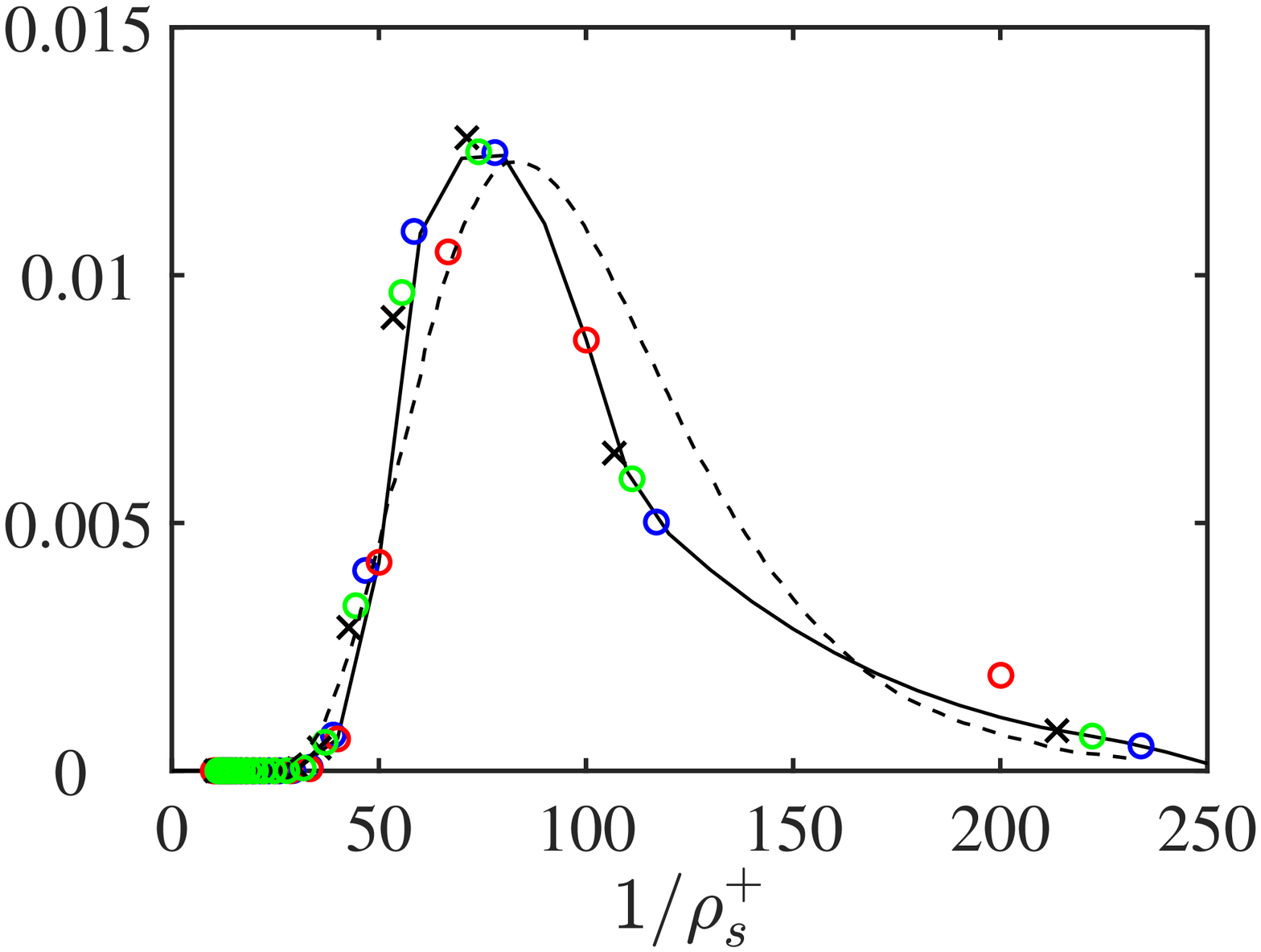}
    \put(-8,70){(\textit{b})}
  \end{overpic}
  }

  \DeclareRobustCommand\mylabela{\tikz[baseline]{\draw[solid, black, thick] (0,0.5ex) -- (0.8,0.5ex);\draw[solid, black, thick] (0.4-0.1,0.5ex-0.6ex) -- (0.4+0.1,0.5ex+0.6ex);\draw[solid, black, thick] (0.4-0.1,0.5ex+0.6ex) -- (0.4+0.1,0.5ex-0.6ex);}}
  \DeclareRobustCommand\mylabelb{\tikz[baseline]{\draw[solid, red, thick] (0,0.5ex) -- (0.8,0.5ex);\draw[solid, red, thick] (0.4-0.1,0.5ex-0.6ex) -- (0.4+0.1,0.5ex+0.6ex);\draw[solid, red, thick] (0.4-0.1,0.5ex+0.6ex) -- (0.4+0.1,0.5ex-0.6ex);}}
  \DeclareRobustCommand\mylabelc{\tikz[baseline]{\draw[solid, blue, thick] (0,0.5ex) -- (0.8,0.5ex);\draw[solid, blue, thick] (0.4-0.1,0.5ex-0.6ex) -- (0.4+0.1,0.5ex+0.6ex);\draw[solid, blue, thick] (0.4-0.1,0.5ex+0.6ex) -- (0.4+0.1,0.5ex-0.6ex);}}
  \DeclareRobustCommand\mylabeld{\tikz[baseline]{\draw[solid, green, thick] (0,0.5ex) -- (0.8,0.5ex);\draw[solid, green, thick] (0.4-0.1,0.5ex-0.6ex) -- (0.4+0.1,0.5ex+0.6ex);\draw[solid, green, thick] (0.4-0.1,0.5ex+0.6ex) -- (0.4+0.1,0.5ex-0.6ex);}}
  \DeclareRobustCommand\mylabele{\tikz[baseline]{\draw[dash pattern={on 2pt off 2pt}, black, thick] (0,0.5ex) -- (0.8,0.5ex);}}

  \DeclareRobustCommand\mylabelf{\tikz[baseline]{\draw[solid, white, thick] (0,0.5ex) -- (0.8,0.5ex);\draw[solid, blue, thick] (0.4,0.5ex) circle (0.1cm);}}
  \DeclareRobustCommand\mylabelg{\tikz[baseline]{\draw[solid, white, thick] (0,0.5ex) -- (0.8,0.5ex);\draw[solid, red, thick] (0.4,0.5ex) circle (0.1cm);}}
  \DeclareRobustCommand\mylabelh{\tikz[baseline]{\draw[solid, white, thick] (0,0.5ex) -- (0.8,0.5ex);\draw[solid, black, thick] (0.4-0.1,0.5ex-0.6ex) -- (0.4+0.1,0.5ex+0.6ex);\draw[solid, black, thick] (0.4-0.1,0.5ex+0.6ex) -- (0.4+0.1,0.5ex-0.6ex);}}
  \DeclareRobustCommand\mylabeli{\tikz[baseline]{\draw[solid, white, thick] (0,0.5ex) -- (0.8,0.5ex);\draw[solid, green, thick] (0.4,0.5ex) circle (0.1cm);}}
  \DeclareRobustCommand\mylabelj{\tikz[baseline]{\draw[dash pattern={on 2pt off 2pt}, black, thick] (0,0.5ex) -- (0.8,0.5ex);}}
  \DeclareRobustCommand\mylabelk{\tikz[baseline]{\draw[solid, black, thick] (0,0.5ex) -- (0.8,0.5ex);}}

\caption{%
(\textit{a}) Probability density function of the streak density at different $\Delta z$ in
case M2000. \mylabela, ${\Delta z}^{+}=214$; \mylabelb, $308$; \mylabelc, $411$; \mylabeld,
$510$; \mylabele, \citet{smet83}.
(\textit{b}) Probability density function of the streak density distributions at different
$\Rey_{\tau}$ and $\Delta z^{+}\approx200$. \mylabelf, L550; \mylabelg, M950; \mylabelh,
M2000; \mylabeli, M4200; \mylabelj, \citet{smet83}; \mylabelk, the best-fitting curve.
}
\label{fig:pdf_rou}
\end{figure}

The spanwise density of streaks $\rho_s$ is the inverse of their average spacing, and is
simply determined from the number, $n_s$, of minima in the observation window,
$\rho_s=n_s/\Delta z$. It follows from the classical estimates of streak spacing that we
can expect $\rho_s^+\approx 0.01$.

\begin{figure}
  \centering
  \subfigure{
  \begin{overpic}
    [scale=0.21]{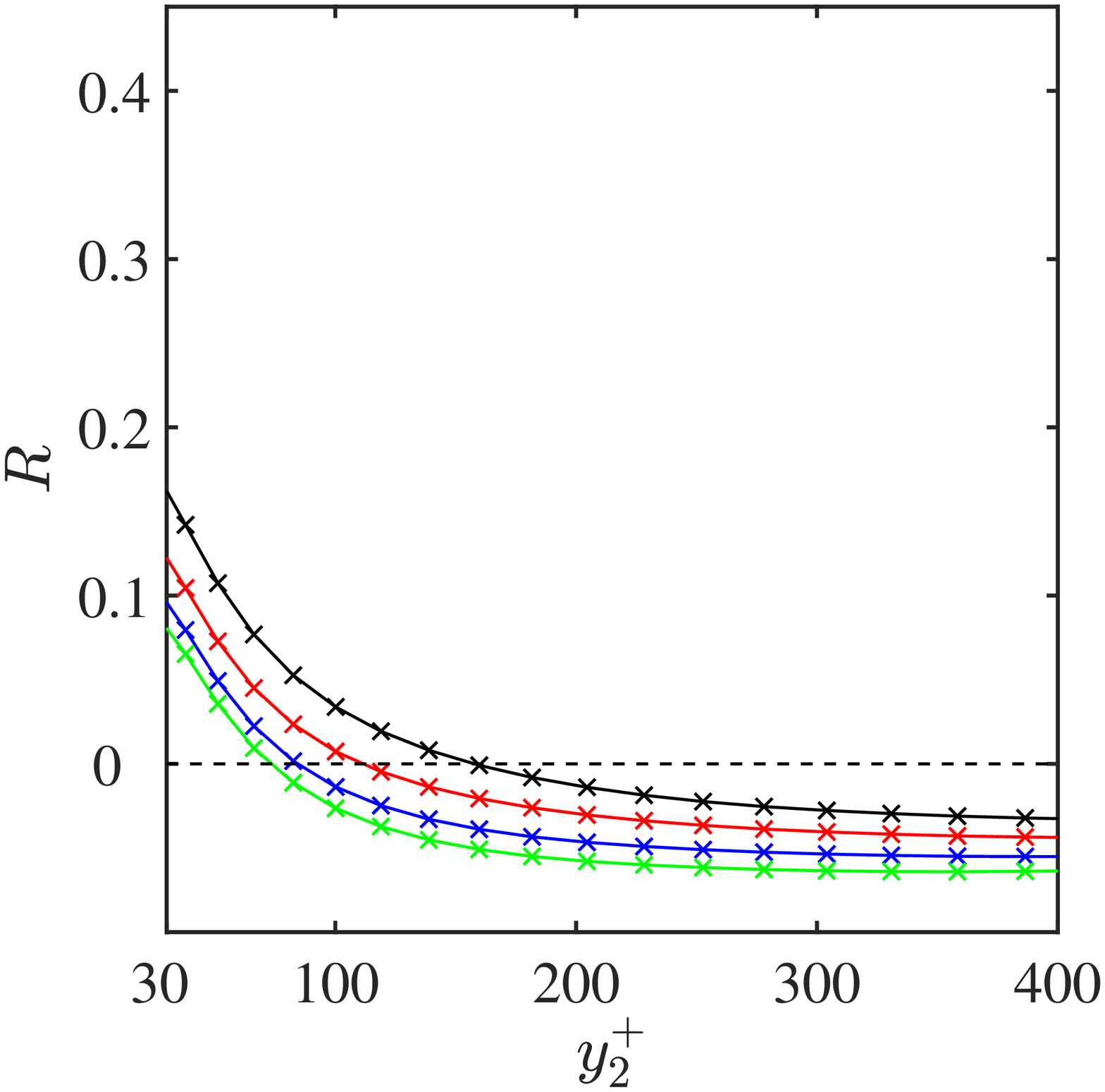}
    \put(-2,93){(\textit{a})}
  \end{overpic}
  \begin{overpic}
    [scale=0.21]{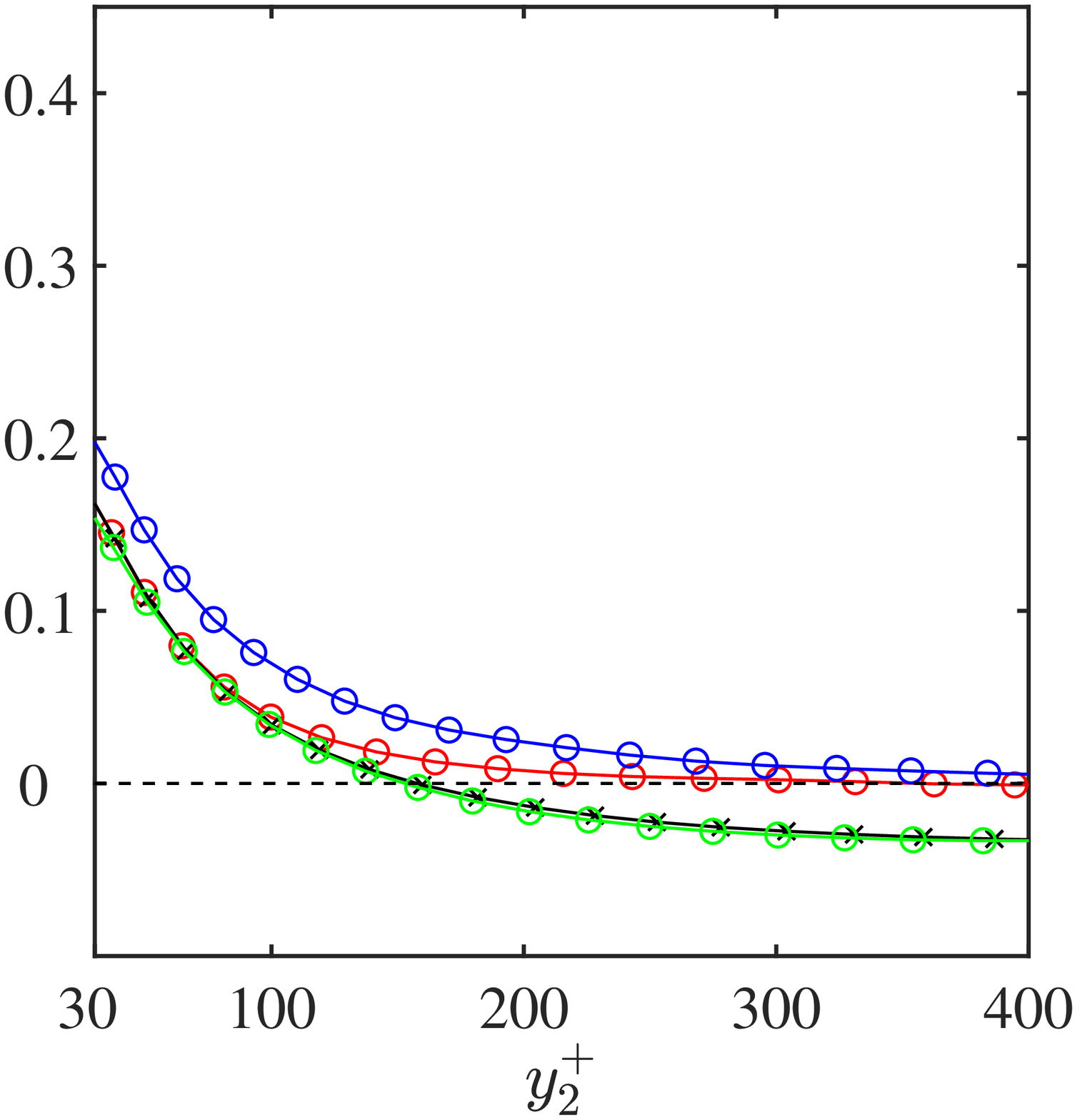}
    \put(-2,93){(\textit{b})}
  \end{overpic}
  \begin{overpic}
    [scale=0.21]{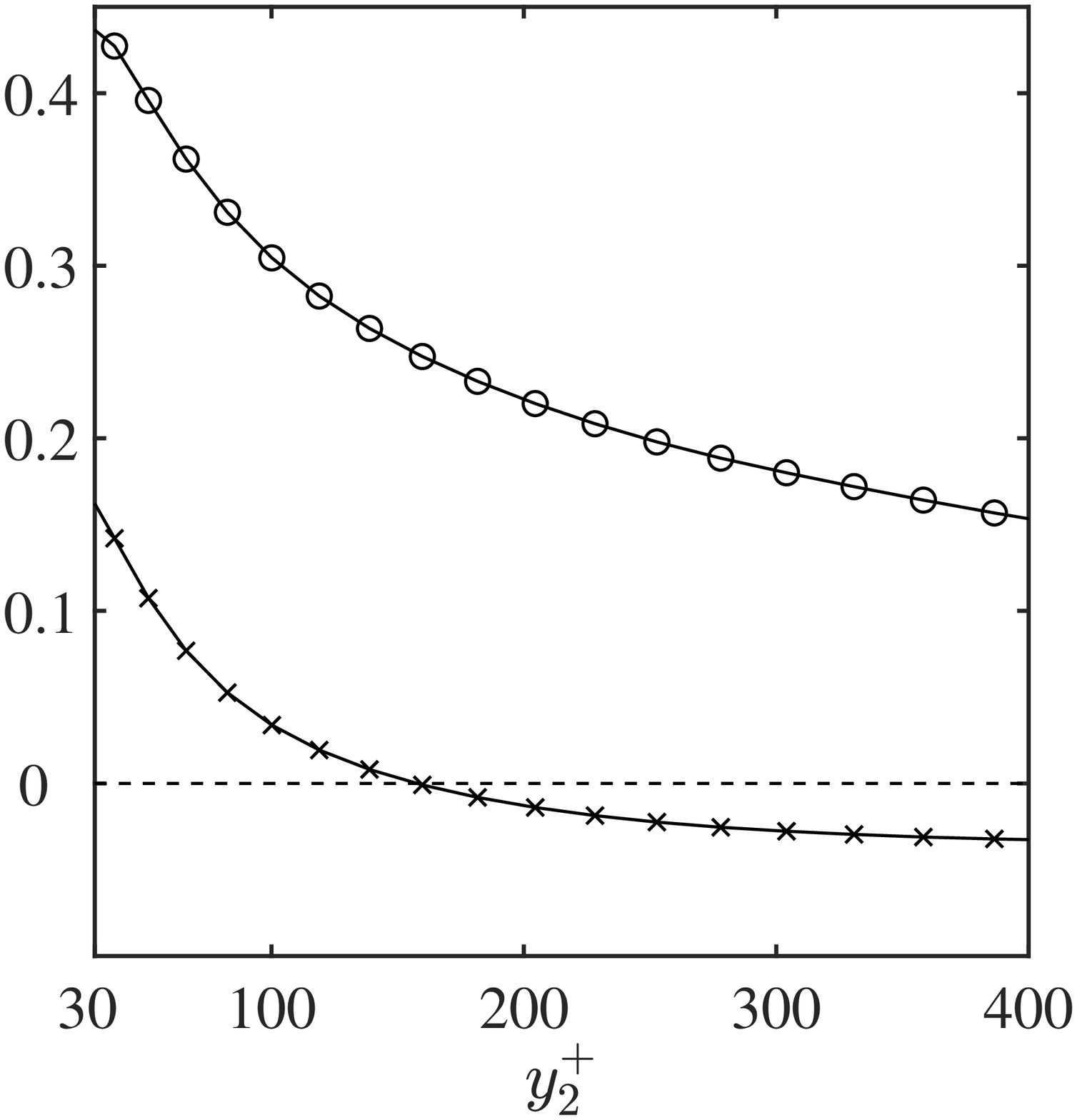}
    \put(-2,93){(\textit{c})}
  \end{overpic}
  }

  \DeclareRobustCommand\mylabela{\tikz[baseline]{\draw[solid, black, thick] (0,0.5ex) -- (0.8,0.5ex);\draw[solid, black, thick] (0.4-0.1,0.5ex-0.6ex) -- (0.4+0.1,0.5ex+0.6ex);\draw[solid, black, thick] (0.4-0.1,0.5ex+0.6ex) -- (0.4+0.1,0.5ex-0.6ex);}}
  \DeclareRobustCommand\mylabelb{\tikz[baseline]{\draw[solid, red, thick] (0,0.5ex) -- (0.8,0.5ex);\draw[solid, red, thick] (0.4-0.1,0.5ex-0.6ex) -- (0.4+0.1,0.5ex+0.6ex);\draw[solid, red, thick] (0.4-0.1,0.5ex+0.6ex) -- (0.4+0.1,0.5ex-0.6ex);}}
  \DeclareRobustCommand\mylabelc{\tikz[baseline]{\draw[solid, blue, thick] (0,0.5ex) -- (0.8,0.5ex);\draw[solid, blue, thick] (0.4-0.1,0.5ex-0.6ex) -- (0.4+0.1,0.5ex+0.6ex);\draw[solid, blue, thick] (0.4-0.1,0.5ex+0.6ex) -- (0.4+0.1,0.5ex-0.6ex);}}
  \DeclareRobustCommand\mylabeld{\tikz[baseline]{\draw[solid, green, thick] (0,0.5ex) -- (0.8,0.5ex);\draw[solid, green, thick] (0.4-0.1,0.5ex-0.6ex) -- (0.4+0.1,0.5ex+0.6ex);\draw[solid, green, thick] (0.4-0.1,0.5ex+0.6ex) -- (0.4+0.1,0.5ex-0.6ex);}}
  \DeclareRobustCommand\mylabele{\tikz[baseline]{\draw[dash pattern={on 2pt off 2pt}, black, thick] (0,0.5ex) -- (0.8,0.5ex);}}

  \DeclareRobustCommand\mylabelf{\tikz[baseline]{\draw[solid, blue, thick] (0,0.5ex) -- (0.8,0.5ex);\draw[solid, blue, thick] (0.4,0.5ex) circle (0.1cm);}}
  \DeclareRobustCommand\mylabelg{\tikz[baseline]{\draw[solid, red, thick] (0,0.5ex) -- (0.8,0.5ex);\draw[solid, red, thick] (0.4,0.5ex) circle (0.1cm);}}
  \DeclareRobustCommand\mylabelh{\tikz[baseline]{\draw[solid, black, thick] (0,0.5ex) -- (0.8,0.5ex);\draw[solid, black, thick] (0.4-0.1,0.5ex-0.6ex) -- (0.4+0.1,0.5ex+0.6ex);\draw[solid, black, thick] (0.4-0.1,0.5ex+0.6ex) -- (0.4+0.1,0.5ex-0.6ex);}}
  \DeclareRobustCommand\mylabeli{\tikz[baseline]{\draw[solid, green, thick] (0,0.5ex) -- (0.8,0.5ex);\draw[solid, green, thick] (0.4,0.5ex) circle (0.1cm);}}
  \DeclareRobustCommand\mylabelj{\tikz[baseline]{\draw[dash pattern={on 2pt off 2pt}, black, thick] (0,0.5ex) -- (0.8,0.5ex);}}

  \DeclareRobustCommand\mylabels{\tikz[baseline]{\draw[solid, black, thick] (0,0.5ex) -- (0.8,0.5ex);\draw[solid, black, thick] (0.4,0.5ex) circle (0.1cm);}}
  \DeclareRobustCommand\mylabelt{\tikz[baseline]{\draw[solid, black, thick] (0,0.5ex) -- (0.8,0.5ex);\draw[solid, black, thick] (0.4-0.1,0.5ex-0.6ex) -- (0.4+0.1,0.5ex+0.6ex);\draw[solid, black, thick] (0.4-0.1,0.5ex+0.6ex) -- (0.4+0.1,0.5ex-0.6ex);}}
\caption{%
(\textit{a}) Correlation $R(\rho_{s},\widetilde{v})$ at different $\Delta z$ in case M2000.
\mylabela, ${\Delta z}^{+}=214$; \mylabelb, $308$; \mylabelc, $411$; \mylabeld, $510$.
(\textit{b}) As in (\aaa) for different $\Rey_{\tau}$ and $\Delta z^{+}\approx200$.
\mylabelf, L550; \mylabelg, M950; \mylabelh, M2000; \mylabeli, M4200.
(\textit{c}) Correlation for $\Delta z^{+}=214$ in case M2000. \mylabels, with condition
(\ref{eq:1}); \mylabelt, without condition (\ref{eq:1}).
}
\label{fig:R_before_utau}
\end{figure}

Distributions of the streak spacing are given in figure \ref{fig:pdf_rou}, compared with the
results of \citet{smet83} in a turbulent boundary layer at $\Rey_{\tau}\approx 724$. Figure
\ref{fig:pdf_rou}(\aaa) shows the effect on the spacing of the size of the detection window.
As expected, the distribution becomes more concentrated for wider windows, but the mode of
the distribution, $\ell_{zu}^{+}\approx 70$, stays constant and is only slightly narrower
than the result of \citet{smet83}, $\ell_{zu}^{+}\approx 85$. Considering that these authors
detect streaks from the accumulation of hydrogen bubbles at $y^+=5$, and determine visually
the distance among neighbouring streaks, the discrepancy can probably be attributed to the
different definitions of what constitutes a streak. Even so, the PDF obtained with our
narrowest window, $\Delta z^+\approx 200$ is fairly close to that in \citet{smet83}, and
this window will be used in the following. Figure \ref{fig:pdf_rou}(\bbb) shows that the
density distribution is essentially independent of the Reynolds number, also in agreement
with \citet{smet83}.

The correlation coefficient between $\rho_{s}$ and $\widetilde{v}$ is defined as
\begin{equation}
R(\rho_{s},\widetilde{v})=\frac{\sum(\rho_{s}-\overline{\rho_{s}})\widetilde{v}}{\left[\sum(\rho_{s}-\overline{\rho_{s}})^{2}\sum\widetilde{v}^{2}\right]^{\frac{1}{2}}},
\label{eq:10}
\end{equation}
where the $\overline{\rho_{s}^{+}}\approx 0.01$ denotes the averaged streak density.
The averaged wall-normal velocity vanishes from continuity, and the summation extends over all
the samples in table \ref{tab:n_sample}.
The streak density $\rho_{s}$ is a local quantity depending on the spanwise locations, while the correlations between the local $\rho_{s}$ and $\widetilde{v}$ could capture the possible streak accumulation.
If the streaks really accumulated in the up-washing
regions of $\widetilde{v}$ and drained away from the down-washing ones, the correlation
$R(\rho_{s},\widetilde{v})$ would be strongly positive. However, figure
\ref{fig:R_before_utau}(\aaa,\bbb) show that the correlation is always smaller than $0.2$,
and even turns negative at large $y_2$, contradicting the assumption of the top-down model of
\citet{toh2005interaction}. This failure can be traced to our rejection of
condition (\ref{eq:1}), which was used by \citet{toh2005interaction} when identifying
streaks. This criterion has very little influence on the drift statistics in
\S\ref{sec:drift}, but figure \ref{fig:R_before_utau}(\ccc), as well as figure \ref{fig:zitaM950new}, shows that it strongly affects
the results for the streak density. The density is clearly correlated with the outer
large-scale circulations when using (\ref{eq:1}), because that condition tends to reject
streaks underneath large-scale high-speed regions, and to retain the roots of the low-speed
ones. But, when only the local minimum condition (\ref{eq:2}) is used, the accumulation
becomes weaker or negative.

Another effect that may influence the correlation $R(\rho_{s},\widetilde{v})$ is the
modulation of the local wall shear by the outer large-scale motions. A negative
$\widetilde{v}$ brings high speed fluid from the outer flow towards the wall, increasing the
local shear and, in effect, decreasing the local viscous length. If the streak generation
process is in equilibrium with this shear, the local streak spacing should decrease and the
streak density should increase. This would tend to counteract the hypothesised local
decrease in density due to divergence, and decrease the inner-outer correlation. If the
modulation is strong enough, it may overcome the correlation completely, and even become
negative.
The effect is similar to the LSM-induced local wall-shear fluctuations in \citet{abe2004very} and the large-scale modulations in \citet{mathis2009large}, which were also shown in \citet{jim12_arfm} and \cite{zha:chern:16} to be mostly reducible to scaling by the local friction velocity, $(u_\tau)_{local}=(\nu\, \partial\widetilde{u}/\partial y)^{\frac{1}{2}}$.

\begin{figure}
  \centering
  \subfigure{
  \begin{overpic}
    [scale=0.32]{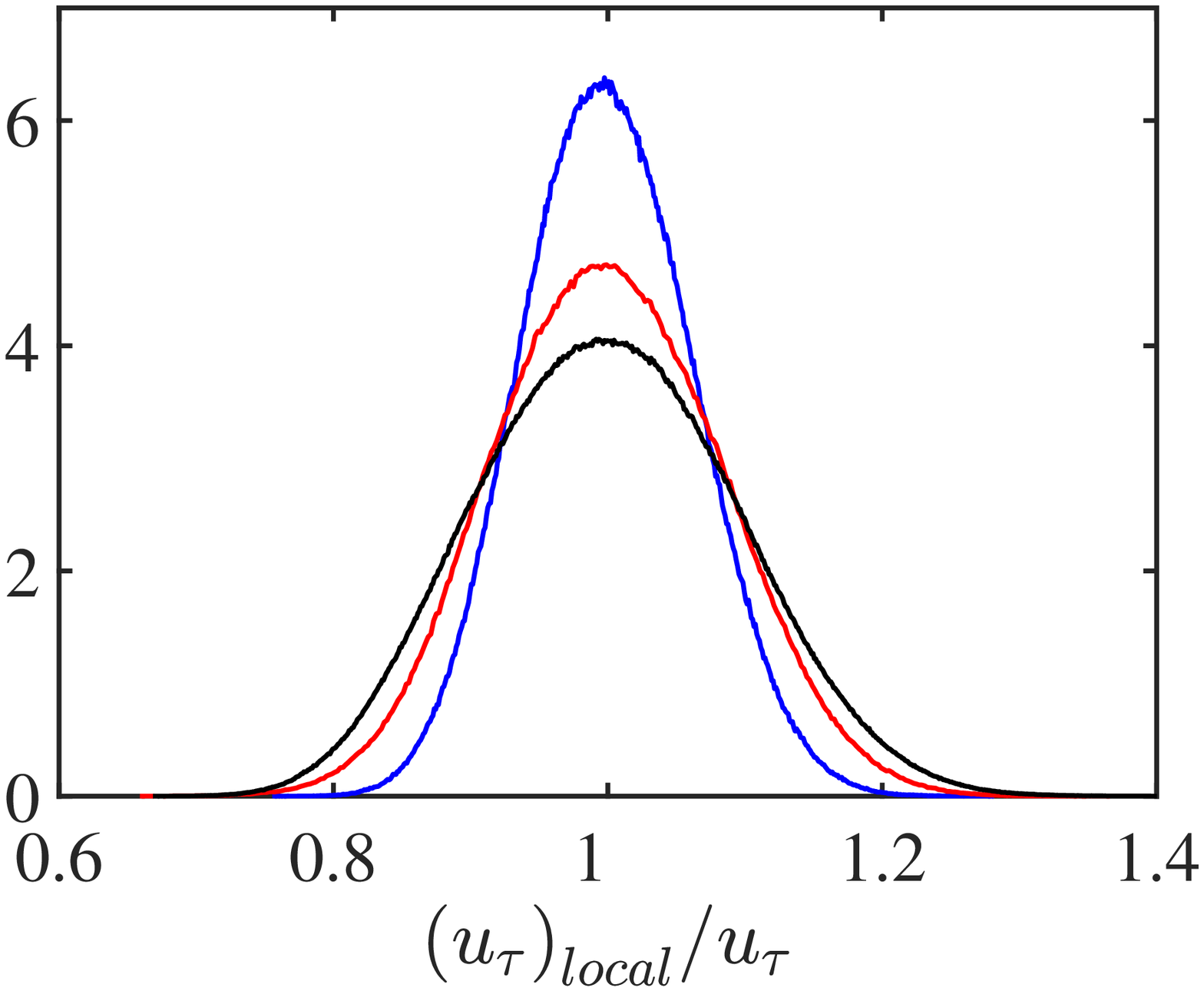}
    \put(2,70){(\textit{a})}
  \end{overpic}
  \begin{overpic}
    [scale=0.32]{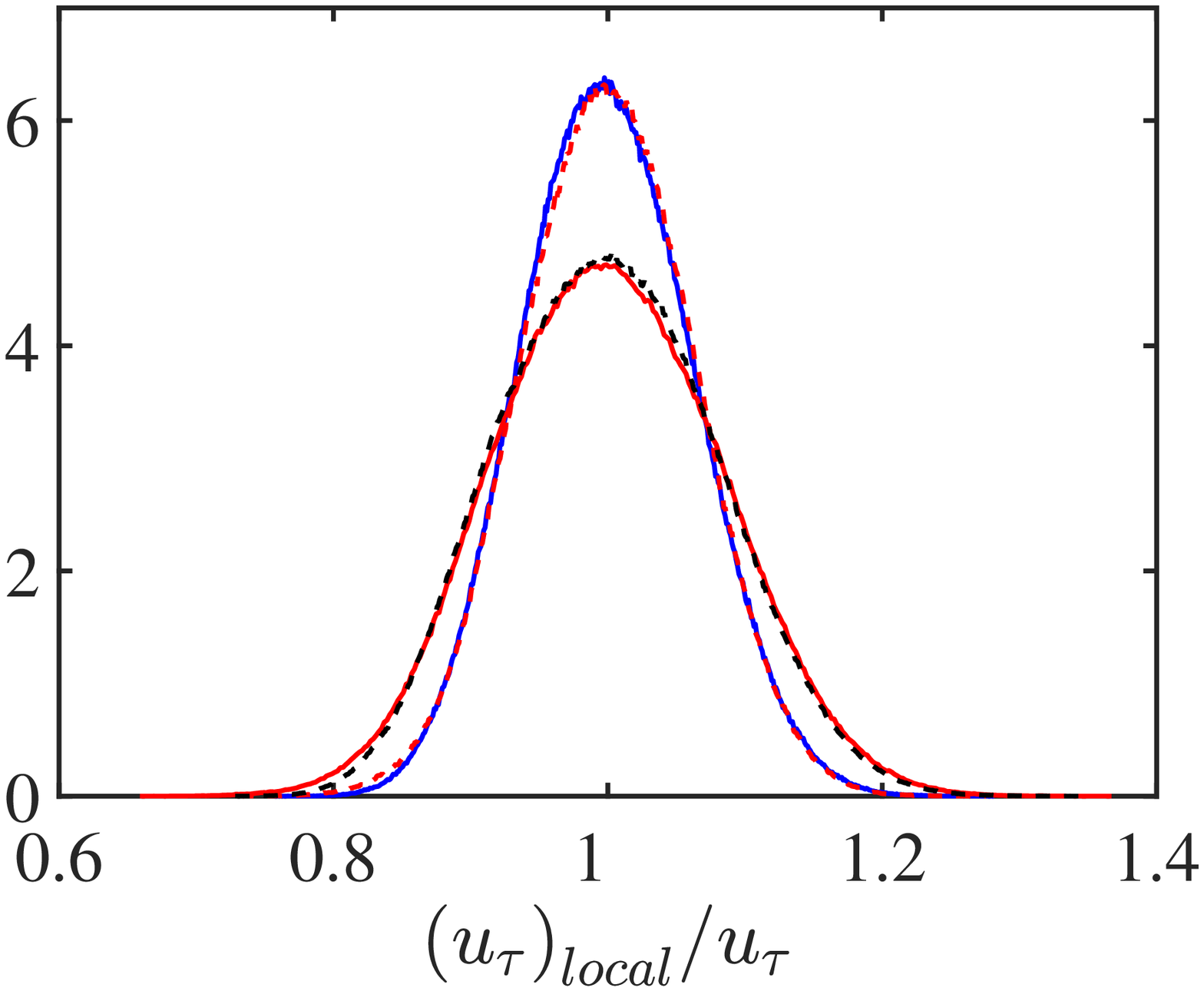}
    \put(2,70){(\textit{b})}
  \end{overpic}
  }

  \DeclareRobustCommand\mylabela{\tikz[baseline]{\draw[solid, blue, thick] (0,0.5ex) -- (0.8,0.5ex);}}
  \DeclareRobustCommand\mylabelb{\tikz[baseline]{\draw[solid, red, thick] (0,0.5ex) -- (0.8,0.5ex);}}
  \DeclareRobustCommand\mylabelc{\tikz[baseline]{\draw[solid, black, thick] (0,0.5ex) -- (0.8,0.5ex);}}
  \DeclareRobustCommand\mylabeld{\tikz[baseline]{\draw[dash pattern={on 2pt off 2pt}, red, thick] (0,0.5ex) -- (0.8,0.5ex);}}

  \DeclareRobustCommand\mylabele{\tikz[baseline]{\draw[dash pattern={on 2pt off 2pt}, black, thick] (0,0.5ex) -- (0.8,0.5ex);}}
  \DeclareRobustCommand\mylabelf{\tikz[baseline]{\draw[solid, green, thick] (0,0.5ex) -- (0.8,0.5ex);}}

  \caption{
Probability density function of $(u_\tau)_{local}/u_\tau$ at different $\Rey_{\tau}$, where
$(u_\tau)_{local}$ is the local friction velocity computed over each window, and $u_\tau$ is
its global average. (\textit{a}) ${\Delta z}^{+}\approx200$. \mylabela, L550; \mylabelb,
M950; \mylabelc, M2000.
(\textit{b}) ${\Delta z}/h\approx0.4$, \mylabela, L550; \mylabeld, M950. ${\Delta
z}/h\approx0.2$, \mylabelb, M950; \mylabele, M2000.
  }
\label{fig:pdfutau}
\end{figure}

Figure \ref{fig:pdfutau} shows that the effect is not trivial, and confirms that it is a consequence
of the interaction with the outer flow. Figure \ref{fig:pdfutau}(\aaa) presents the PDF of
$(u_\tau)_{local}$ for three Reynolds numbers, computed over a spanwise averaging window
whose size scales in wall units. The variation is of the order of $\pm 20\%$, and gets wider
as the Reynolds number increases, in agreement with the increase in energy from the wider
range of scales of the velocity fluctuations. If the effect were local to the near-wall
layer, it would be difficult to explain this Reynolds number dependence. In fact, figure
\ref{fig:pdfutau}(\bbb) shows that, when the averaging window is scaled in outer units, the
variation of $(u_\tau)_{local}$ depends on the size of the window, but not on the Reynolds
number.


\begin{figure}
  \centering
  \subfigure{
  \begin{overpic}
    [scale=0.21]{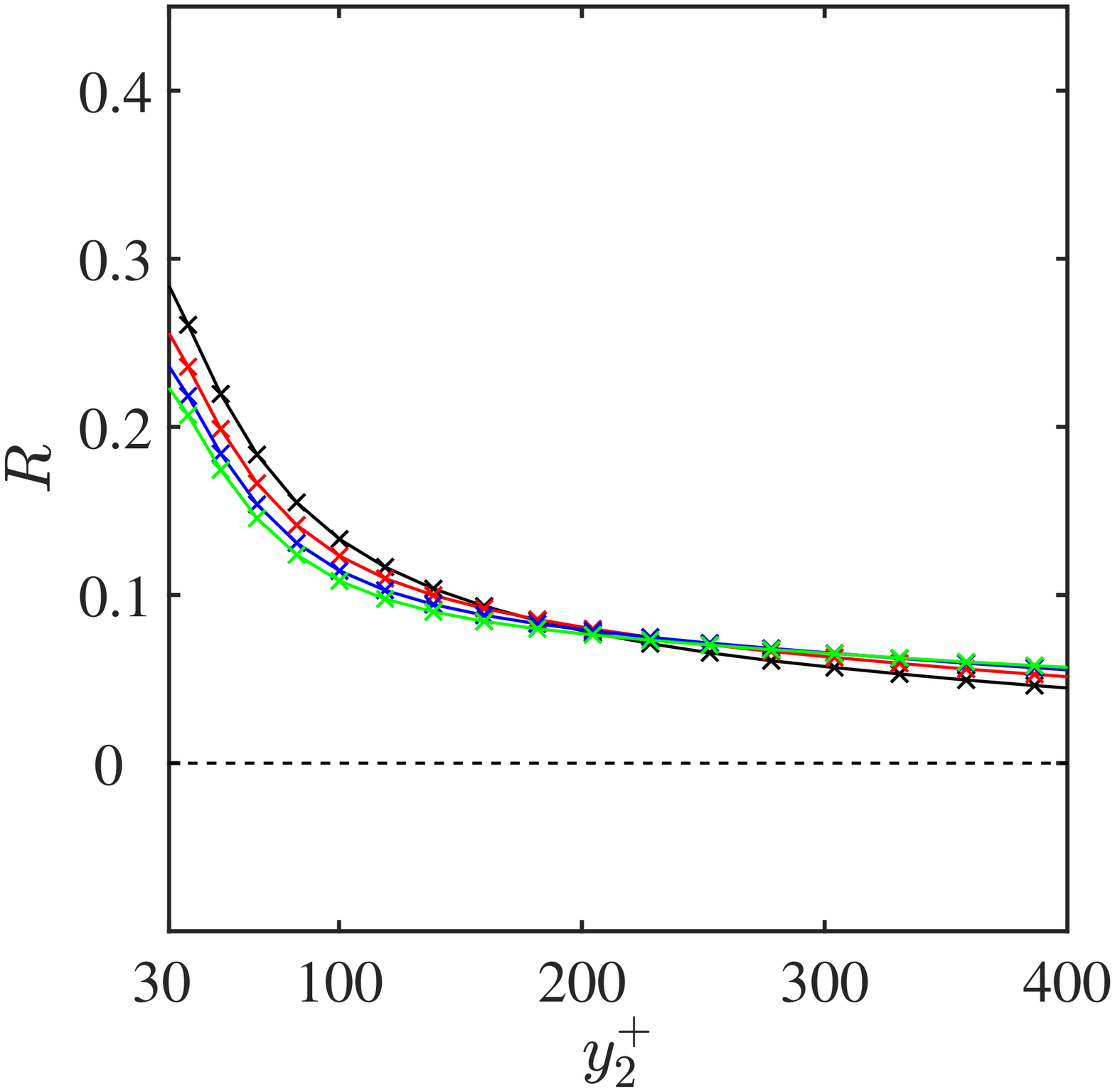}
    \put(-2,93){(\textit{a})}
  \end{overpic}
  \begin{overpic}
    [scale=0.21]{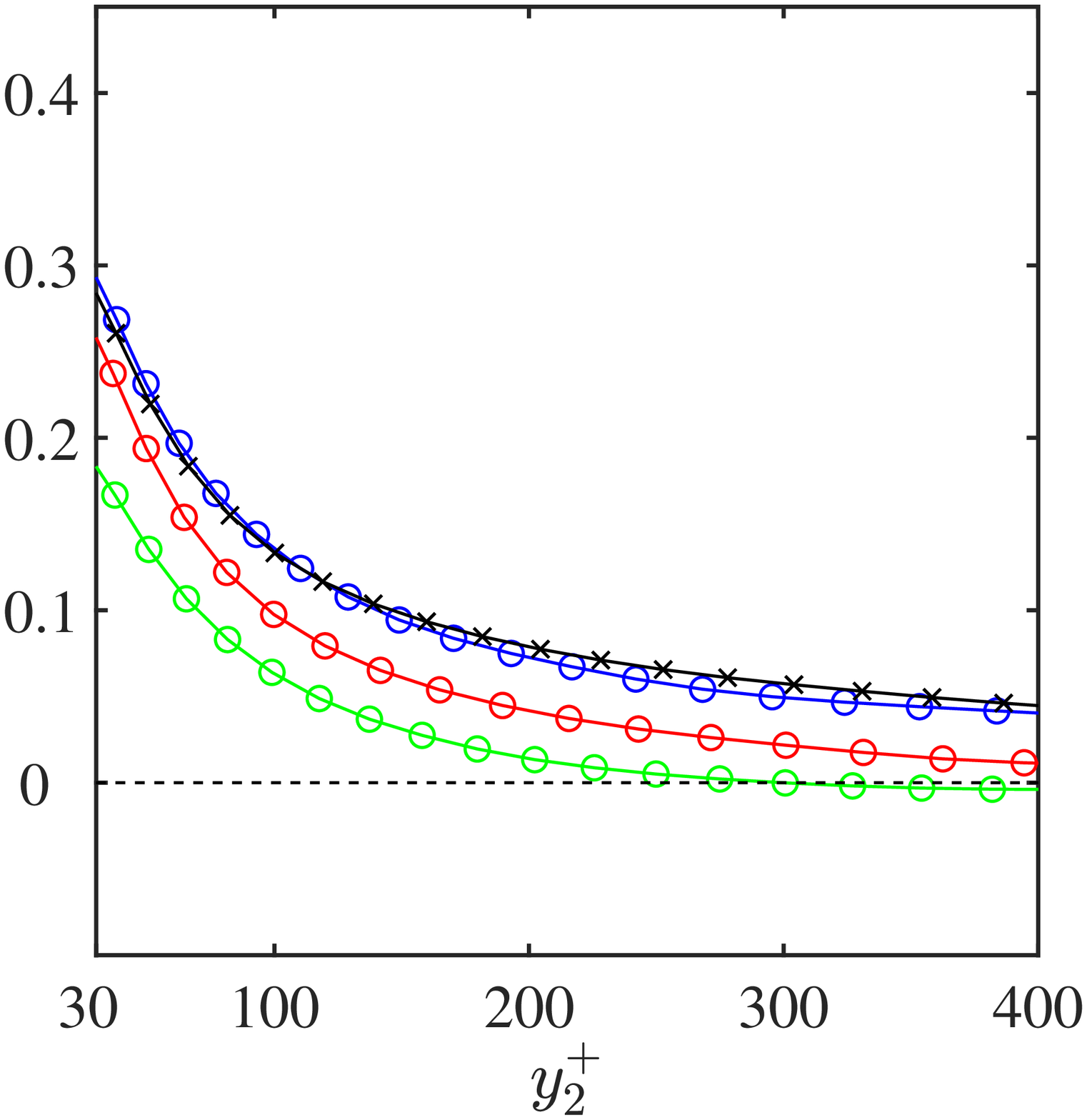}
    \put(-2,93){(\textit{b})}
  \end{overpic}
  }

  \DeclareRobustCommand\mylabela{\tikz[baseline]{\draw[solid, black, thick] (0,0.5ex) -- (0.8,0.5ex);\draw[solid, black, thick] (0.4-0.1,0.5ex-0.6ex) -- (0.4+0.1,0.5ex+0.6ex);\draw[solid, black, thick] (0.4-0.1,0.5ex+0.6ex) -- (0.4+0.1,0.5ex-0.6ex);}}
  \DeclareRobustCommand\mylabelb{\tikz[baseline]{\draw[solid, red, thick] (0,0.5ex) -- (0.8,0.5ex);\draw[solid, red, thick] (0.4-0.1,0.5ex-0.6ex) -- (0.4+0.1,0.5ex+0.6ex);\draw[solid, red, thick] (0.4-0.1,0.5ex+0.6ex) -- (0.4+0.1,0.5ex-0.6ex);}}
  \DeclareRobustCommand\mylabelc{\tikz[baseline]{\draw[solid, blue, thick] (0,0.5ex) -- (0.8,0.5ex);\draw[solid, blue, thick] (0.4-0.1,0.5ex-0.6ex) -- (0.4+0.1,0.5ex+0.6ex);\draw[solid, blue, thick] (0.4-0.1,0.5ex+0.6ex) -- (0.4+0.1,0.5ex-0.6ex);}}
  \DeclareRobustCommand\mylabeld{\tikz[baseline]{\draw[solid, green, thick] (0,0.5ex) -- (0.8,0.5ex);\draw[solid, green, thick] (0.4-0.1,0.5ex-0.6ex) -- (0.4+0.1,0.5ex+0.6ex);\draw[solid, green, thick] (0.4-0.1,0.5ex+0.6ex) -- (0.4+0.1,0.5ex-0.6ex);}}
  \DeclareRobustCommand\mylabele{\tikz[baseline]{\draw[dash pattern={on 2pt off 2pt}, black, thick] (0,0.5ex) -- (0.8,0.5ex);}}

  \DeclareRobustCommand\mylabelf{\tikz[baseline]{\draw[solid, blue, thick] (0,0.5ex) -- (0.8,0.5ex);\draw[solid, blue, thick] (0.4,0.5ex) circle (0.1cm);}}
  \DeclareRobustCommand\mylabelg{\tikz[baseline]{\draw[solid, red, thick] (0,0.5ex) -- (0.8,0.5ex);\draw[solid, red, thick] (0.4,0.5ex) circle (0.1cm);}}
  \DeclareRobustCommand\mylabelh{\tikz[baseline]{\draw[solid, black, thick] (0,0.5ex) -- (0.8,0.5ex);\draw[solid, black, thick] (0.4-0.1,0.5ex-0.6ex) -- (0.4+0.1,0.5ex+0.6ex);\draw[solid, black, thick] (0.4-0.1,0.5ex+0.6ex) -- (0.4+0.1,0.5ex-0.6ex);}}
  \DeclareRobustCommand\mylabeli{\tikz[baseline]{\draw[solid, green, thick] (0,0.5ex) -- (0.8,0.5ex);\draw[solid, green, thick] (0.4,0.5ex) circle (0.1cm);}}
  \DeclareRobustCommand\mylabelj{\tikz[baseline]{\draw[dash pattern={on 2pt off 2pt}, black, thick] (0,0.5ex) -- (0.8,0.5ex);}}

  \caption{
  Correlation $R(\rho_{s}^{+},\widetilde{v}^{+})$ scaled in local wall units.
  (\textit{a}) At different window size in case M2000.
  \mylabela, ${\Delta z}^{+}\approx214$; \mylabelb, $308$; \mylabelc, $411$; \mylabeld, $510$.
  (\textit{b}) At different $\Rey_{\tau}$ when $\Delta z^{+}\approx200$.
  \mylabelf, L550; \mylabelg, M950; \mylabelh, M2000; \mylabeli, M4200.
}
\label{fig:R_after_utau}
\end{figure}

Finally, figure \ref{fig:R_after_utau} shows that, when both $\rho_{s}^{+}$ and
$\widetilde{v}^{+}$ are scaled in the local wall units, the effect of the averaging window
on $R(\rho_{s}^{+},\widetilde{v}^{+})$ largely disappears, but the correlation still decays
with $y_2$. It becomes negligible above $y_2^+\approx 150$ but, contrary to figure
\ref{fig:R_before_utau}(\aaa,\bbb), it does not become negative at higher $y_2$. It is interesting
that the limit for this decay scales in wall units, suggesting that the reason has more to
do with the dynamics of the buffer layer than with the outer flow.

What needs to be explained is why streaks drift spanwise but do not accumulate, and the
simplest explanation is their short lifetime, which is $t^{+}=O(500)$ instead of the
$t^{+}=1000-3000$ in \citet{toh2005interaction}. Since $w_s^+=O(1)$, their maximum drift is
thus only $\delta z^+ =300-500$ (see figure \ref{fig:zitaM950new}\aaa), and the outer
structures can only couple with the streak density over widths of this order. It follows
from figure \ref{fig:kiRet} that this corresponds to $y_2^+\approx 100$, and that the
accumulation hypothesised in \citet{toh2005interaction} is a wall-layer effect that should
become increasingly less relevant as the Reynolds number increases.

\section{Bottom-up influence on the LSM generation and preservation}\label{sec:bottomup}

In order to examine the possibility of bottom-up influence during the LSM generation
process, two further numerical experiments are performed. In the first one, the flow in the
channel is initialized with a laminar velocity profile near the upper wall and a turbulent
velocity profile near the lower wall, as in \citet{schoppa2002coherent}. The initial
perturbations, added below $y^{+}=50$ in the lower half of the channel, are constructed by filtering the flow in case W535 to retain the velocity fluctuations with $\lambda_{z}^{+}<230$. Since
experience shows that early flow adjustments kill part of these perturbations, they are
amplified by a moderate factor (approximately 1.3) before being added to the flow, so that
the near-wall Reynolds stress is similar to the fully-developed one after the initial decay.
This ensures that the channel transitions to turbulence, and is robust to amplification
factors up to two.
Part of the adjustment is the imposition of continuity by the first
simulation step, which is not initially satisfied near the upper boundary of the added
streaks $(y^{+}=50)$, but the effect is minor. The code and computational parameters are as
in W535 (see table \ref{tab:cp}), with the Reynolds number fixed to $\Rey=U_{m}h/\nu=9800$,
which corresponds to $\Rey_{\tau}\approx 535$ near the lower wall.

%
%

\begin{figure}
    \centering
    \begin{overpic}
    [scale=0.28]{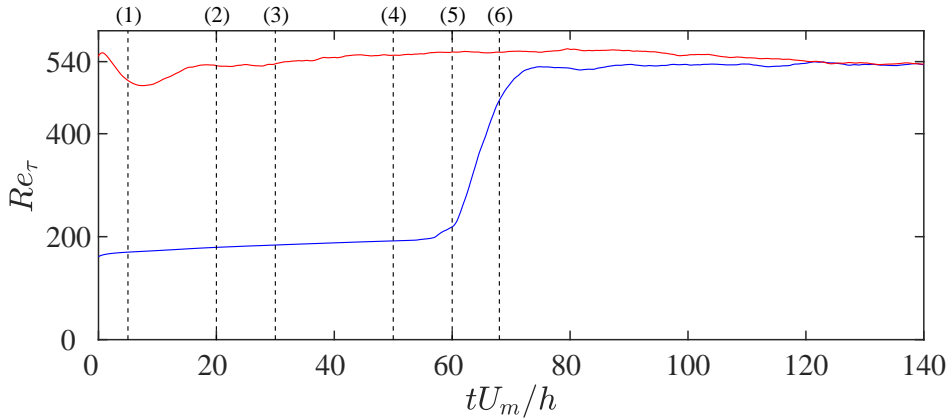}
    \end{overpic}

    \DeclareRobustCommand\mylabela{\tikz[baseline]{\draw[solid, blue, thick] (0,0.5ex) -- (0.8,0.5ex);}}
    \DeclareRobustCommand\mylabelb{\tikz[baseline]{\draw[solid, red, thick] (0,0.5ex) -- (0.8,0.5ex);}}

    \caption{Time history of the local $\Rey_{\tau}$ on the upper (\mylabela) and lower (\mylabelb) walls. The dashed vertical lines (1) to (6) mark the corresponding time instants in figure \ref{fig:structure1} and figure \ref{fig:structure2}. (1) $t=5$, (2) $t=20$, (3) $t=30$, (4) $t=50$, (5) $t=60$, (6) $t=68$.}
    \label{fig:u_tau}
\end{figure}

\begin{figure}
  \centering
  \subfigure{
  \begin{overpic}
    [scale=0.26]{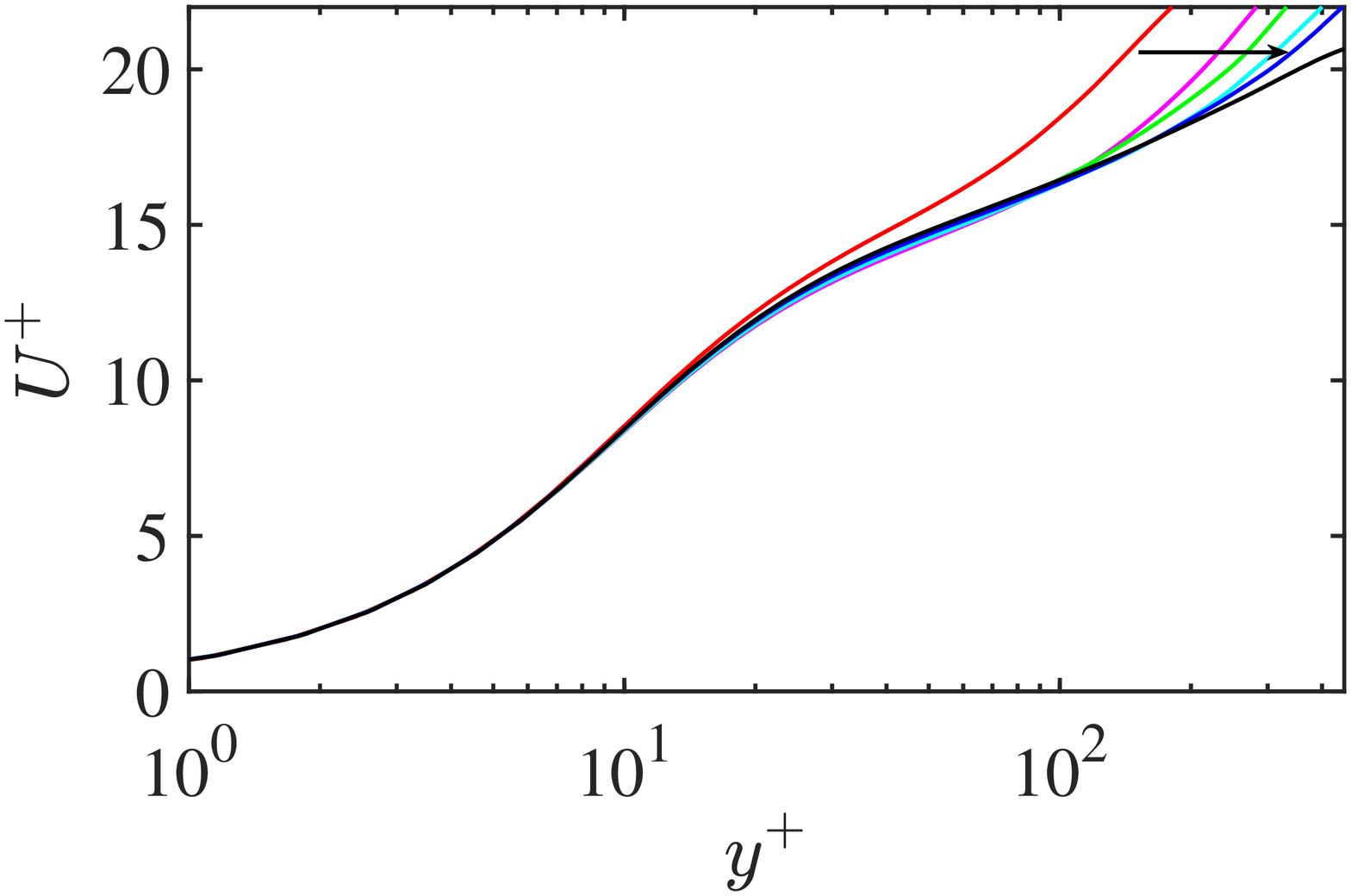}
    \put(0,60){(\textit{a})}
  \end{overpic}
  \begin{overpic}
    [scale=0.26]{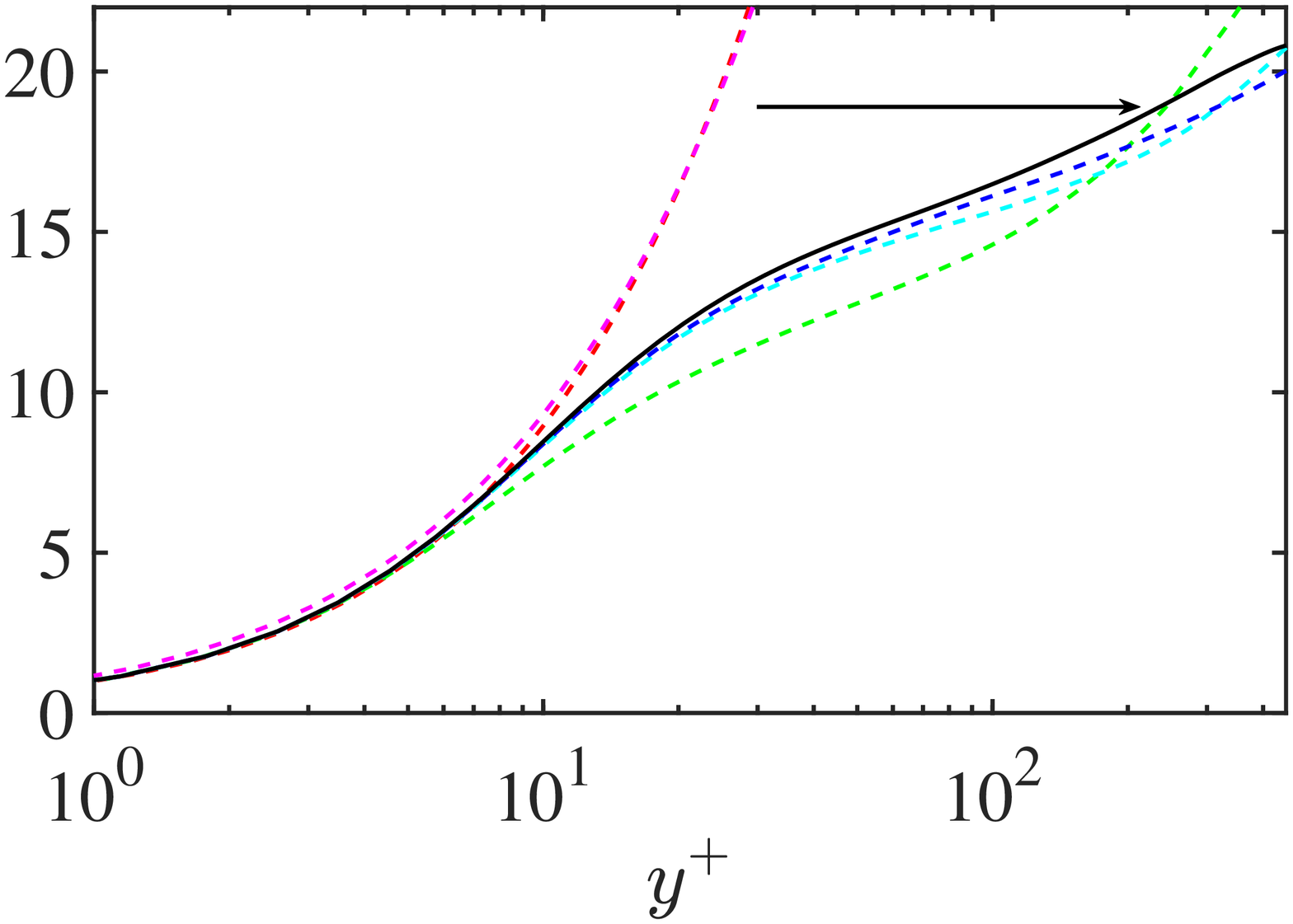}
    \put(1,60){(\textit{b})}
  \end{overpic}
  }

  \DeclareRobustCommand\mylabela{\tikz[baseline]{\draw[solid, red, thick] (0,0.5ex) -- (0.8,0.5ex);}}
  \DeclareRobustCommand\mylabelb{\tikz[baseline]{\draw[solid, m00, thick] (0,0.5ex) -- (0.8,0.5ex);}}
  \DeclareRobustCommand\mylabelc{\tikz[baseline]{\draw[solid, green, thick] (0,0.5ex) -- (0.8,0.5ex);}}
  \DeclareRobustCommand\mylabeld{\tikz[baseline]{\draw[solid, c00, thick] (0,0.5ex) -- (0.8,0.5ex);}}
  \DeclareRobustCommand\mylabele{\tikz[baseline]{\draw[solid, blue, thick] (0,0.5ex) -- (0.8,0.5ex);}}

  \DeclareRobustCommand\mylabelf{\tikz[baseline]{\draw[dash pattern={on 2pt off 2pt}, red, thick] (0,0.5ex) -- (0.8,0.5ex);}}
  \DeclareRobustCommand\mylabelg{\tikz[baseline]{\draw[dash pattern={on 2pt off 2pt}, m00, thick] (0,0.5ex) -- (0.8,0.5ex);}}
  \DeclareRobustCommand\mylabelh{\tikz[baseline]{\draw[dash pattern={on 2pt off 2pt}, green, thick] (0,0.5ex) -- (0.8,0.5ex);}}
  \DeclareRobustCommand\mylabeli{\tikz[baseline]{\draw[dash pattern={on 2pt off 2pt}, c00, thick] (0,0.5ex) -- (0.8,0.5ex);}}
  \DeclareRobustCommand\mylabelj{\tikz[baseline]{\draw[dash pattern={on 2pt off 2pt}, blue, thick] (0,0.5ex) -- (0.8,0.5ex);}}

  \DeclareRobustCommand\mylabelk{\tikz[baseline]{\draw[solid, black, thick] (0,0.5ex) -- (0.8,0.5ex);}}

  \caption{Time evolution of the mean velocity profile, $U^{+}$, scaled in local wall units.
  (\textit{a}) In the lower half-channel from $t=10$ to $t=50$:
  \mylabela, $t=10$; \mylabelb, $t=20$; \mylabelc, $t=30$; \mylabeld, $t=40$; \mylabele, $t=50$.
  And (\textit{b}) in the upper half-channel, from $t=50$ to $t=90$:
  \mylabelf, $t=50$; \mylabelg, $t=60$; \mylabelh, $t=70$; \mylabeli, $t=80$; \mylabelj, $t=90$.
  The black lines (\mylabelk) denote the profile in case W535. The arrows indicate increasing time. }
\label{fig:U_mean}
\end{figure}

Before detailed quantitative analysis, it is helpful to have an overview of the development
of the flow. Figure \ref{fig:u_tau} shows the time history of the local $\Rey_{\tau}$ on the
upper and lower walls. The lower-wall $\Rey_{\tau}$ initially decreases a bit,
and then increases until $t=20$, gradually approaching its target value of
$\Rey_{\tau}=535$. On the upper wall, the flow remains laminar until turbulent transition
occurs around $t=60$. The upper $\Rey_{\tau}$ then quickly increases until $t=75$, after which it
approaches its turbulent level, $\Rey_{\tau}=535$. The time evolution of the mean
velocity profile is displayed in figure \ref{fig:U_mean}, non-dimensionalised by the
local wall units at the corresponding time instants. In the lower side, shown in figure
\ref{fig:U_mean}(\textit{a}), the profile overlaps the fully-developed one below
$y^{+}=20$ for $t=10-50$, and the agreement gradually extends to higher positions as time
develops. This suggests that turbulent structures are gradually being constructed at higher
positions as the flow approaches equilibrium. After $t=50$, the mean velocity profile
collapses to the fully-developed flow. In the upper half-channel, shown in figure
\ref{fig:U_mean}(\textit{b}), the profile quickly changes from laminar to turbulent as that
wall transitions at $t=60-70$, as also indicated by the abrupt increase of
$\Rey_{\tau}$ in figure \ref{fig:u_tau}. The flow continues to adjusts after $t=80$, and the
whole channel then evolves to a fully-developed turbulent state.

\begin{figure}
  \centering
  \begin{overpic}
    [scale=0.48]{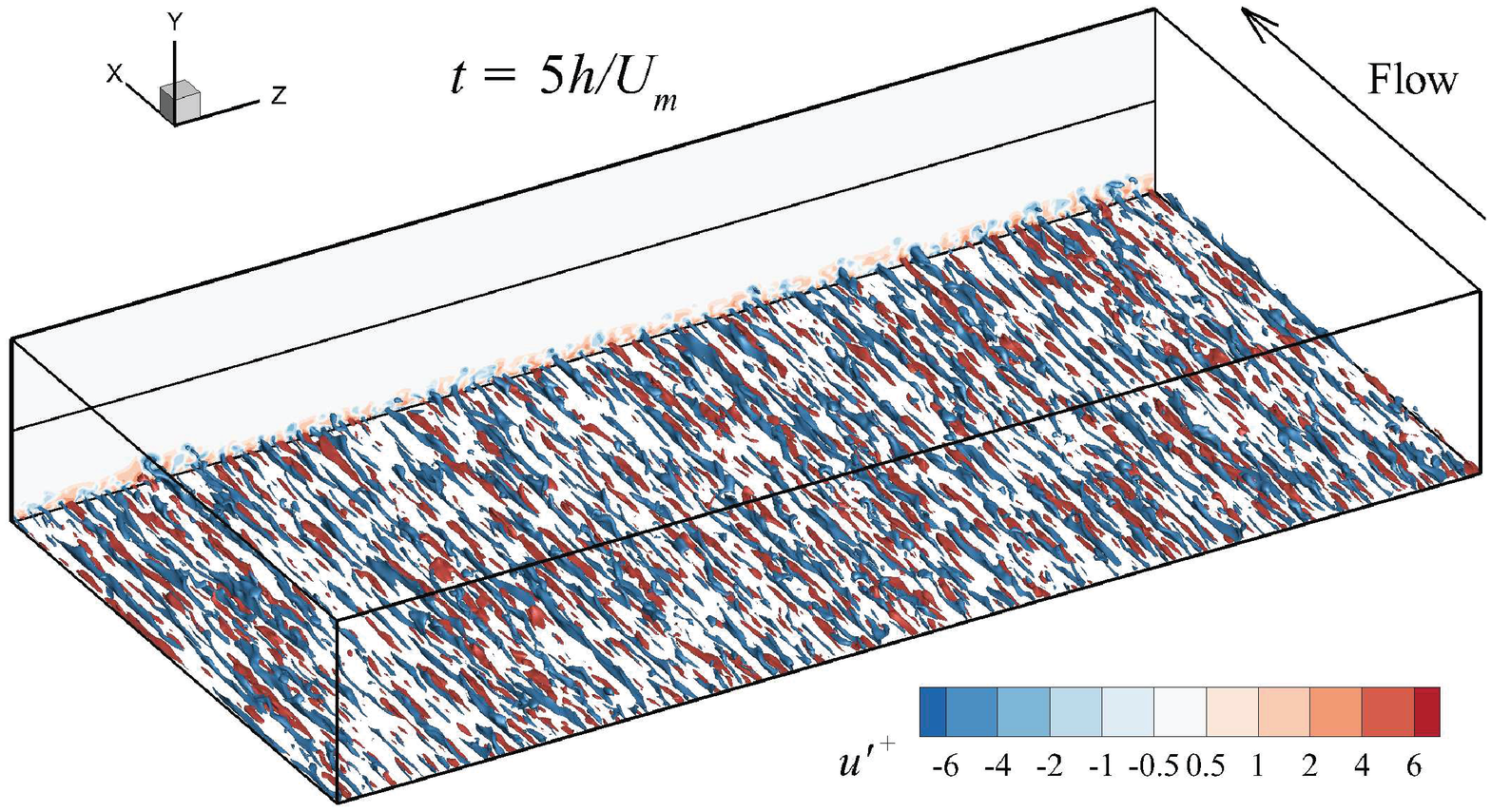}
    \put(-4,50){(\textit{a})}
  \end{overpic}

  \begin{overpic}
    [scale=0.48]{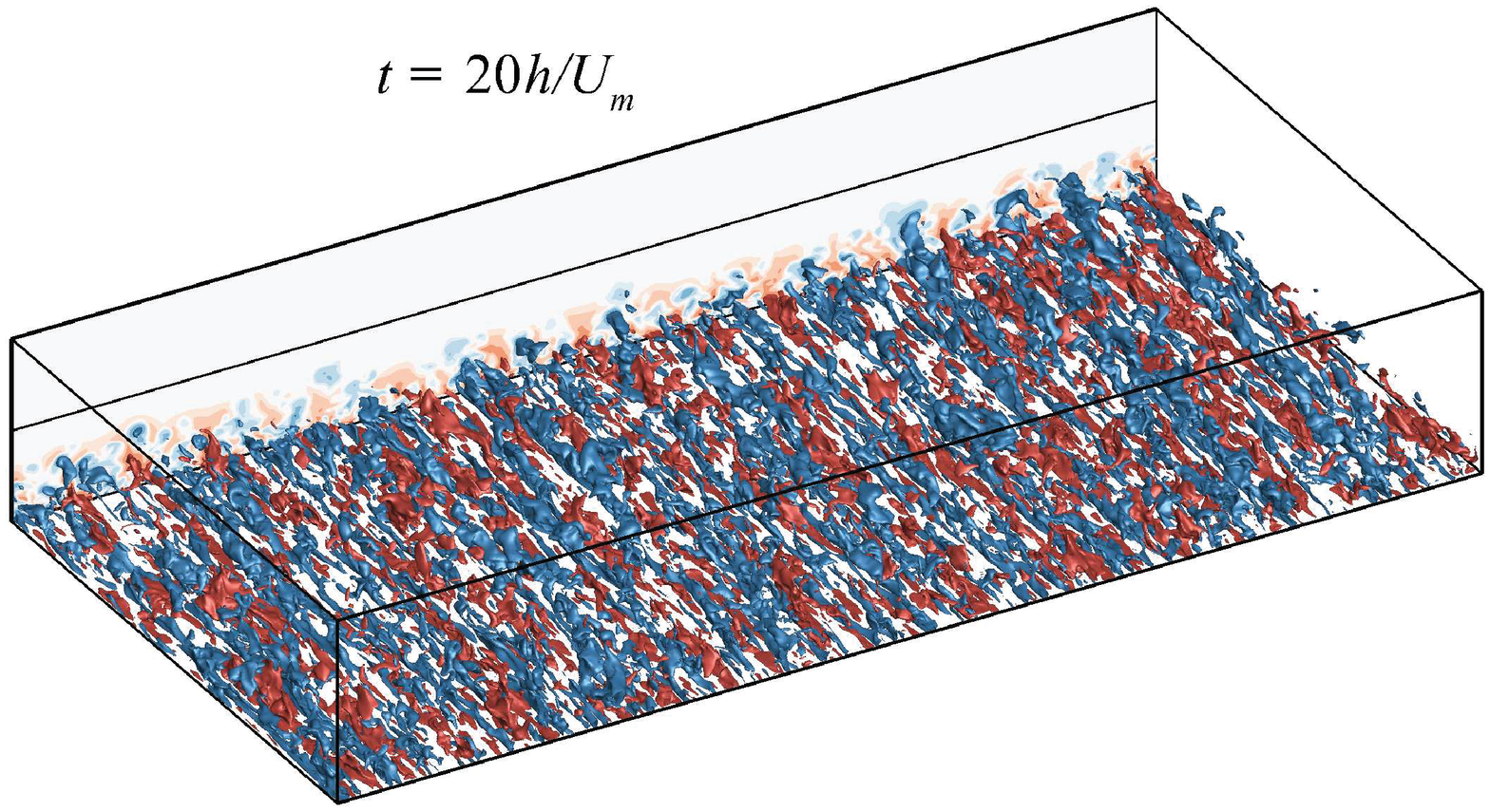}
    \put(-4,50){(\textit{b})}
  \end{overpic}

  \begin{overpic}
    [scale=0.48]{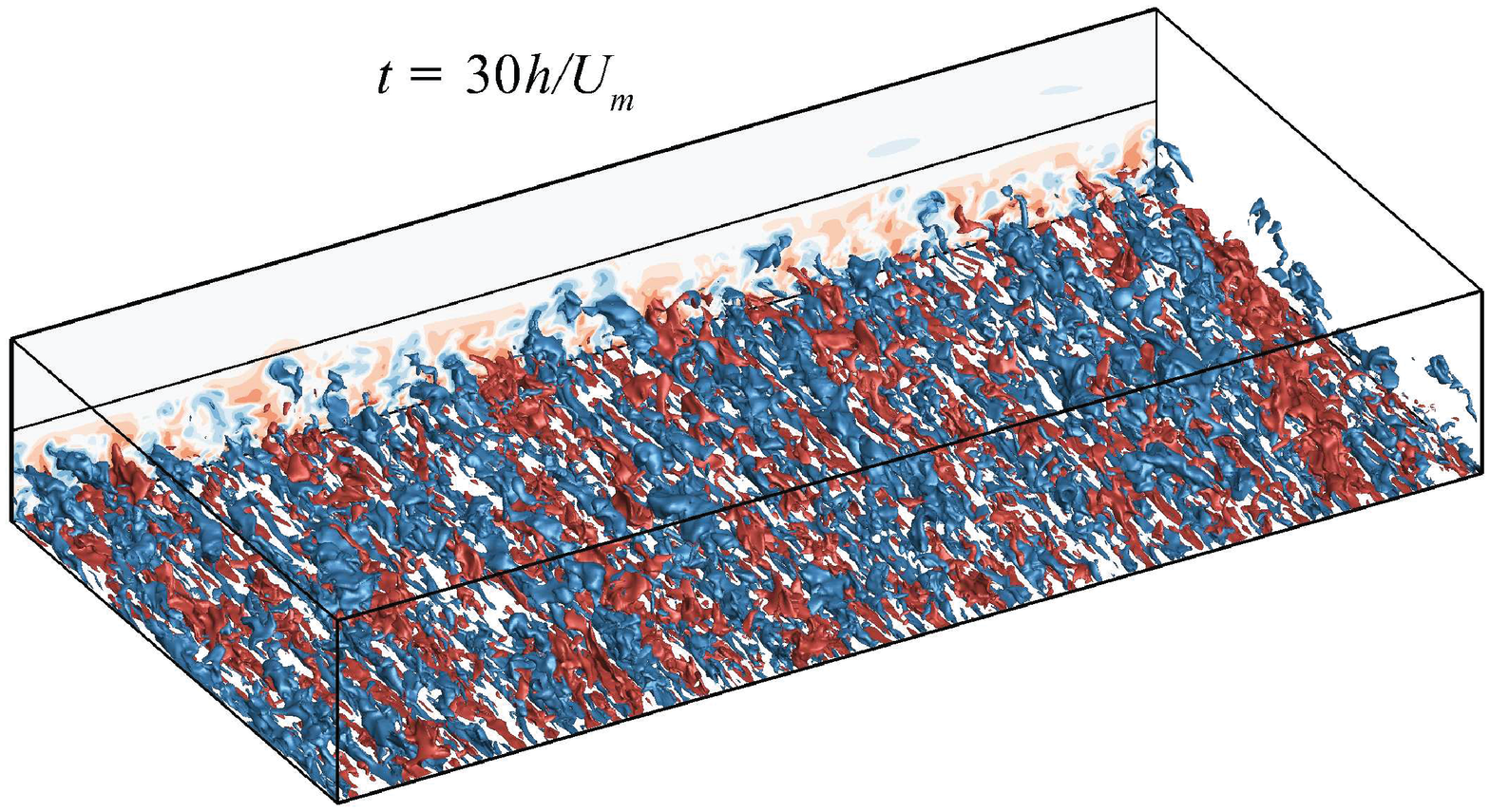}
    \put(-4,50){(\textit{c})}
  \end{overpic}

  \caption{Distributions of $u^{\prime}$ in the lower half-channel at (\textit{a}) $t=5$, (\textit{b}) $t=20$ and (\textit{c}) $t=30$.
  Length in the $x$ direction is ${2\pi}$ and in the $z$ direction is $4\pi$.
  Red: ${u^{\prime}}^{+}=3$, blue: ${u^{\prime}}^{+}=-3$.
  Flow is from bottom-right to top-left.}
\label{fig:structure1}
\end{figure}

We consider in this section the evolution of the flow structures near the lower wall, where
the initial perturbations are added. The evolution of the upper half-channel is closer to classical
bypass transition, and is discussed in appendix \ref{sec:upperside}. The distribution of the
streamwise velocity fluctuations in the lower half-channel is displayed in figure
\ref{fig:structure1} at the three selected times indicated by the lines (1) to (3) in figure
\ref{fig:u_tau}. At $t=5$, the perturbations are confined to the near-wall region,
which is dominated by small-scale velocity streaks. Turbulent fluctuations gradually spread
farther from the wall, in accordance with the evolution of the velocity profile in figure
\ref{fig:U_mean}, and streaks of increasingly larger scales appear at $t=20$ and $t=30$. The
approximate size of the largest streaks at $t=20$ is $\lambda_{z}\sim 0.7h$, and they grow to
$\lambda_{z}\sim h$ at $t=30$. The turbulent fluctuations visually occupy the whole lower
half-channel after that time.

\begin{figure}
    \centering
    \begin{overpic}
    [scale=0.35]{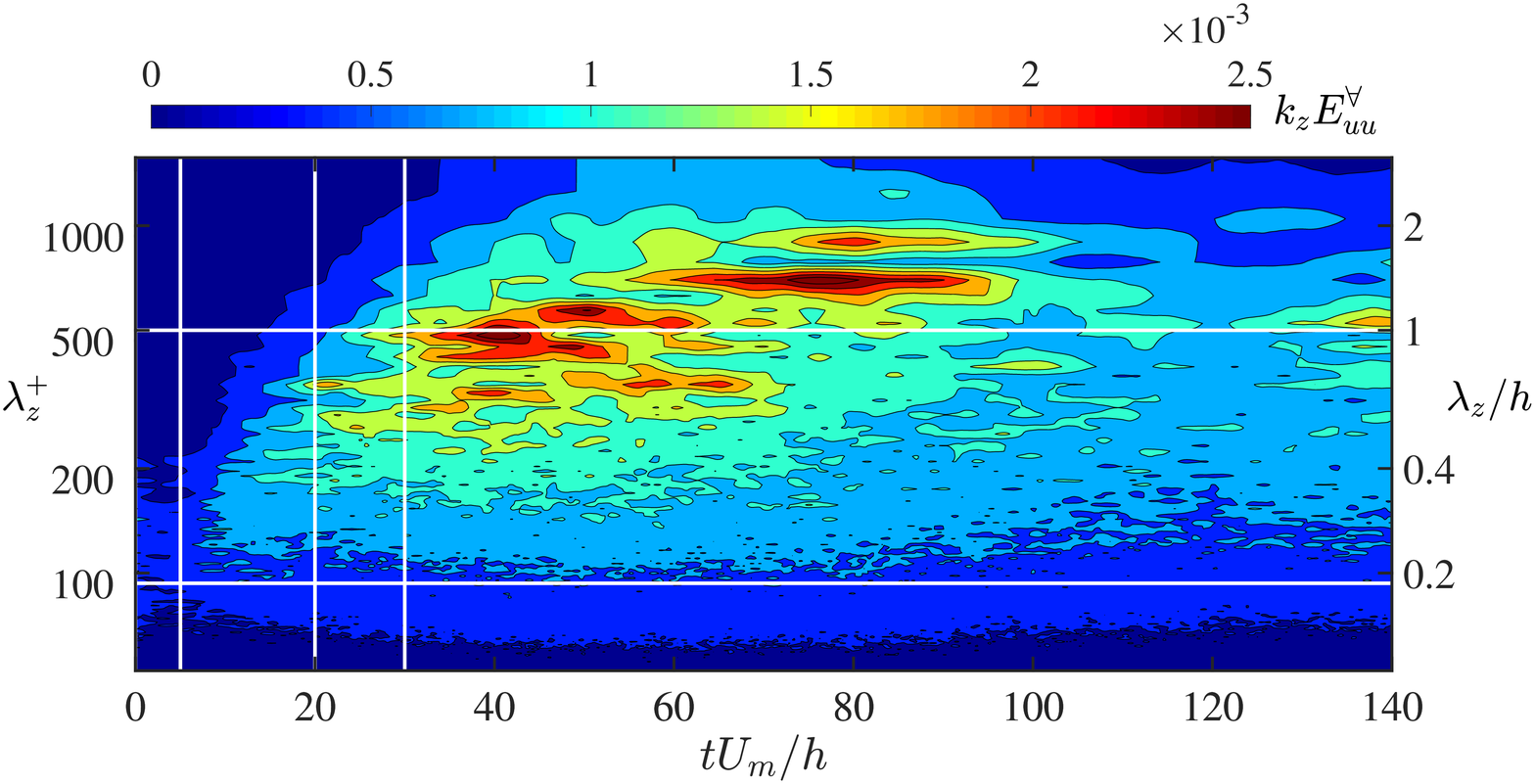}
    \end{overpic}
    \caption{Time evolution of ${k}_{z}E_{uu}^{\forall}(t,\lambda_{z})$ as a function of $\lambda_{z}$ in the lower half-channel. The vertical white lines mark  the time instants $t=5$, $20$ and $30$.}
    \label{fig:energyt}
\end{figure}

The evolution of the streamwise velocity fluctuations towards larger scales can be quantified by the
time-dependent spanwise energy spectrum, defined as
\begin{equation}
  E_{uu}(t,y,k_{z})=\frac{1}{L_{x}}\intop_{0}^{L_{x}}
  (\left|\widehat{u}(t,x,y,k_{z})\right|^{2}+\left|\widehat{u}(t,x,y,-k_{z})\right|^{2})\,\mathrm{d}x.
\end{equation}
The evolution of its integral over the lower half-channel,
$E_{uu}^{\forall}(t,\lambda_{z})$, is shown in figure \ref{fig:energyt} as a function of
$\lambda_{z}$ and of time. The energy is initially concentrated around
$100<\lambda_{z}^{+}<230$, where the initial perturbations have been introduced, and spreads
to wider wavelengths as large-scale motions are gradually generated. The final peak at
$\lambda_{z}\sim h$ develops for $t\gtrsim 30$, in accordance with figure \ref{fig:structure1}.
The numerical experiment evolves to a fully-developed turbulent state after $t=120$, with much lower fluctuations compared to the peaks before $t=100$.

\begin{figure}
    \centering
    \begin{overpic}
    [scale=0.3]{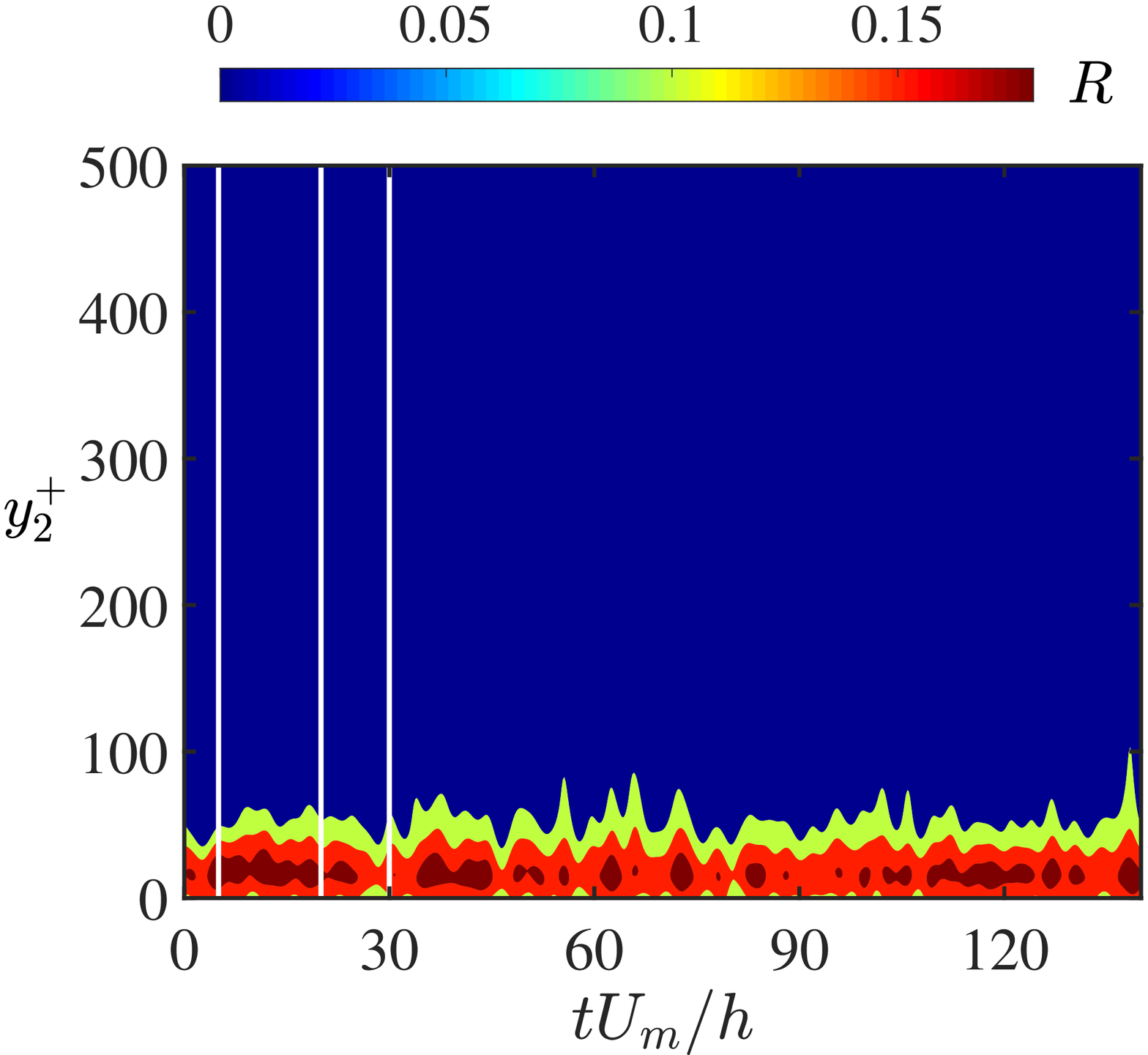}
    \put(-2,82){(\textit{a})}
    \end{overpic}
    \begin{overpic}
    [scale=0.3]{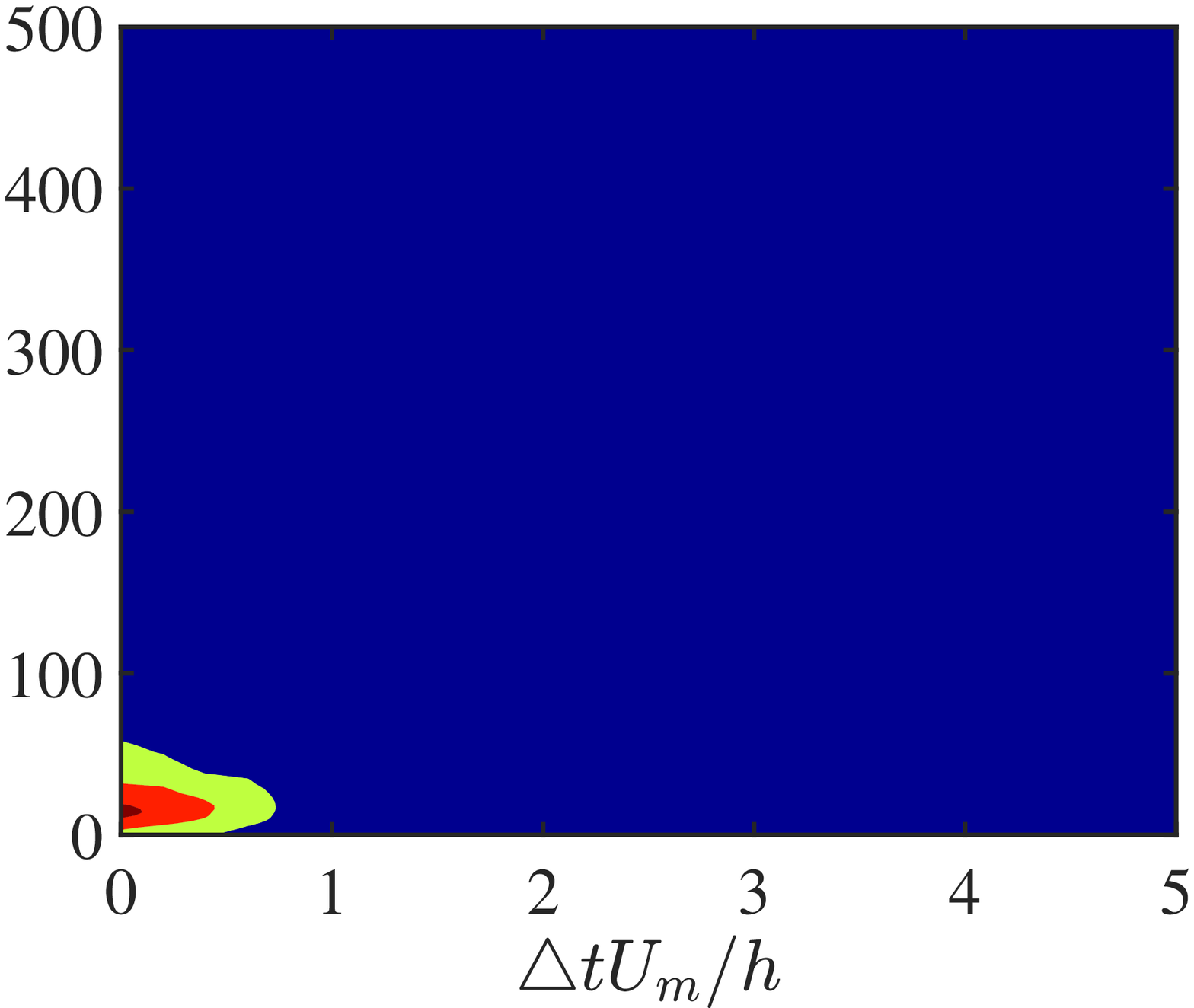}
    \put(0,82){(\textit{b})}
    \end{overpic}
    \caption{(\textit{a}) Time history of the correlation $R(\rho_{s},\widetilde{v})$ with $\Delta z^{+}\approx 200$. Vertical white lines mark the time instants $t=5$, $20$ and $30$.
    (\textit{b}) The correlation $R$ between $\widetilde{v}(x,y_{2},z,t_{0})$ and $\rho_{s}(x-x_{ad},y_{1},z,t_{0}-\Delta t)$ as function of $\Delta t$. $\Delta z^{+}\approx200$. The statistics are compiled by scanning $x$ and $z$. $t_{0}=30h/U_{m}$, and $x_{ad}=(u_{ad}(y_{2})-u_{ad}(y_{1}))\Delta t$ is the streamwise offset due to the different advection velocities.}
    \label{fig:rhost}
\end{figure}

After confirming that the flow is able to generate LSMs with the only input of initially
imposed near-wall streaks, the question of whether streak accumulation is required for this
generation is explored by means of the temporal evolution of the correlation \r{eq:10}
between the near-wall streak density, $\rho_{s}(y_1^+=13)$, and the wall-normal velocity
$\widetilde{v}(y_2)$. Figure \ref{fig:rhost}(\textit{a}) presents the time history of this
correlation coefficient $R(t, y_2)$, with an averaging window $\Delta z^{+}\approx 200$.

The wall-normal locations for which $R>0.1$ are confined below $y_2^+=100$, implying only
local interactions within the buffer layer, and no change can be observed before or after
the LSMs appear at $t \approx 30$. Therefore, the conclusion reached in \S\ref{sec:density}
about the lack of correlation between the streak density and the outer velocity also applies
to the LSM generation process.

The same lack of correlation holds when a delay is introduced in $R$. The correlation
between $\widetilde{v}(y_2, t_0)$ and $\rho_{s}(y_1,t_{0}-\Delta t)$ is shown in figure
\ref{fig:rhost}(\textit{b}) as a function of $\Delta t$. If the outer LSMs were a result of
streak accumulation, as assumed by the bottom-up hypothesis, this correlation should peak
at some finite delay, but it is clear from the figure that the correlation maximum is at
$\Delta t=0$, and that it never reaches above $y^+_2\approx 30$. This again argues against
the hypothesis that the formation of outer scales is due to the previous accumulation of
streaks in the near-wall region.

While this numerical experiment shows that the generation of LSMs does not depend on the
accumulation of the near-wall streaks, it could still be possible that the maintenance of the
LSMs depends on the existence of large-scale nonuniformities near the wall. This is
tested by a second numerical experiment, in which the LSMs below $y^{+}=50$ are artificially
removed after the flow has become fully turbulent in the first experiment.

\begin{figure}
    \centering
    \begin{overpic}
    [scale=0.33]{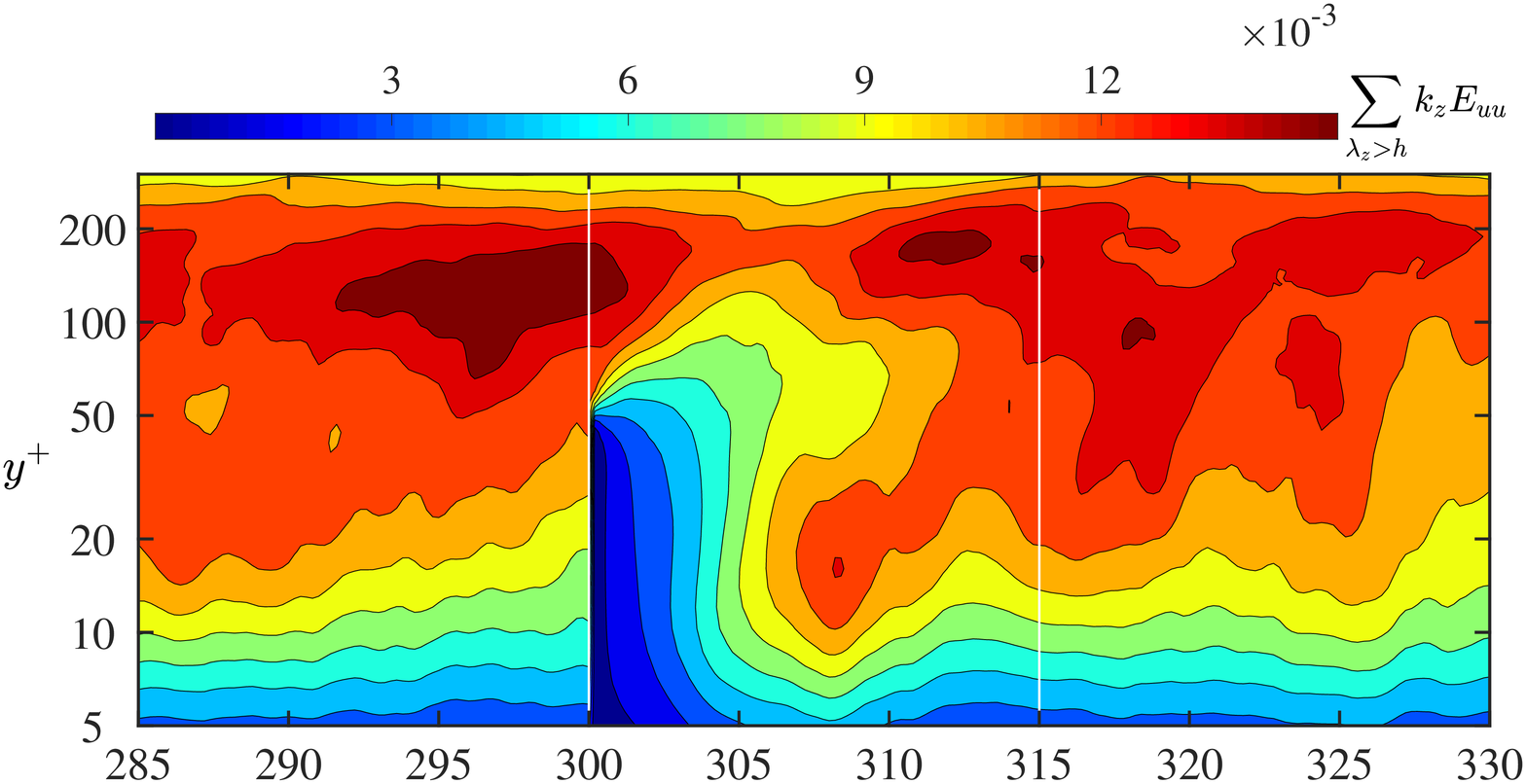}
    \put(3,46){(\textit{a})}
    \end{overpic}

    \begin{overpic}
    [scale=0.33]{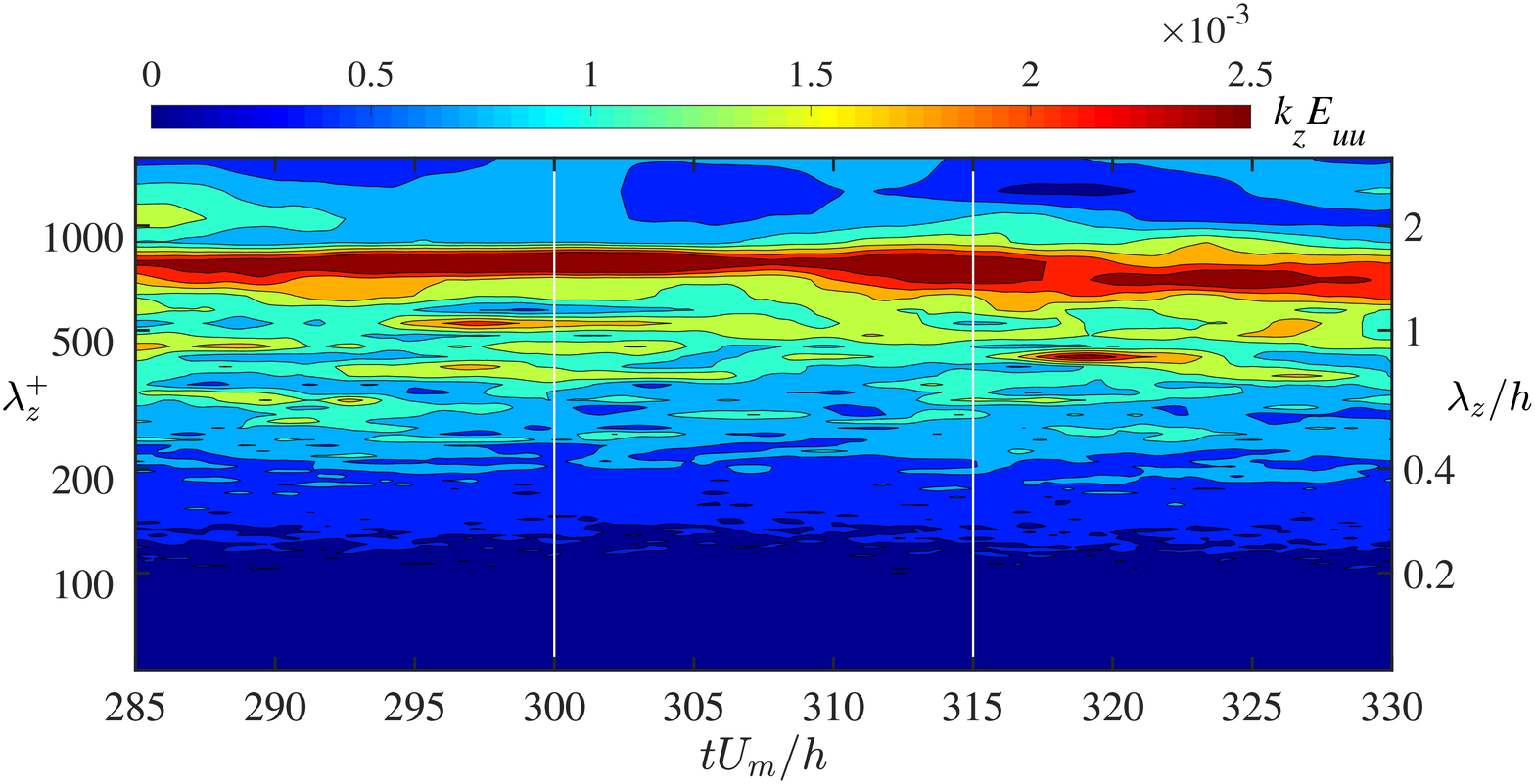}
    \put(3,46){(\textit{b})}
    \end{overpic}

\caption{Time evolution of:
(\textit{a}) $\sum_{\lambda_{z}>h}k_{z}E_{uu}(t,y^{+},\lambda_{z})$ and (\textit{b}) $k_{z}E_{uu}$ at
$y^{+}=200$ in the lower half channel from $t=285$ to $t=330$. The vertical white lines are
$t=300$ and $t=315$.}
    \label{fig:energyty_new}
\end{figure}

Fourier filtering is applied at $t=300$ to the lower side of the channel below $y^{+}=50$,
to remove the fluctuations of the three velocity components with $\lambda_{z}>h/2$. The
subsequent evolution of the streamwise turbulent kinetic energy at scales $\lambda_{z}>h$ is
displayed in figure \ref{fig:energyty_new}(\textit{a}). The influence of the filtering
gradually rises to $y^{+}\approx 100$, while new energy appears in the buffer layer at
$t\approx 305$, with a local maximum near $t=308$, possibly reflecting the top-down
influence of the LSMs remaining in the outer region. The adjustment of the buffer layer
lasts until $t\thickapprox315$, after which $E_{uu}$ recovers its equilibrium distribution.
The time evolution of $k_{z}E_{uu}$ at $y^{+}=200$ during this recovery process is displayed
in figure \ref{fig:energyty_new}(\textit{b}). The energy peak at $\lambda_{z}\approx 1.5h$
evolves smoothly with time, and the influence of the disturbed buffer layer can hardly be
seen.

This result suggests that the removal of LSMs below $y^{+}=50$ has little impact on the
preservation of LSMs in the outer region. Notably, the adjustment time of the buffer layer in
response to the sudden filtering is about ${\Delta t}=15h/U_{m}$ ($\Delta
t^{+}\thickapprox440$), which is of the same order as the average lifetime of LSMs at the
present Reynolds number \citep{lozano2014time}, and should thus be long enough for
the LSMs to decay if their maintenance mechanism had been interrupted. Figure
\ref{fig:energyty_new}(\textit{b}) shows that they do not.

In summary, the twin experiments in this section strongly suggest that the production and
preservation of LSMs in the outer region do not rely on the existence of large-scale
organisation in the near-wall region, in accordance with the results of
\citet{flores2006effect} and others mentioned in the Introduction. They all suggest that the
possible bottom-up influence is not essential for the preservation of the outer
structures.

\section{Summary and conclusions}\label{sec:conc}

The inner-outer co-supporting model of \citet{toh2005interaction}, which had up to now only
been studied in turbulent channels with short simulation boxes, has been examined in
full-sized simulations at low to moderate Reynolds numbers. The model has three stages: the
near-wall streaks drift in the spanwise direction under the influence of the outer large-scale motions
(LSMs); they accumulate in areas of large-scale velocity convergence; and areas of high
streak density lead to the formation of new LSMs.

We have confirmed the first stage by tracking near-wall streaks by a method similar to
particle image velocimetry. The streaks drift spanwise with velocities of the order of $\pm
u_\tau$, and this drift is correlated with the large-scale velocity of the outer
structures. Moreover, the coupling not only happens between the wall and
the largest LSMs. Structures centred at distance $y$ from the wall in the logarithmic layer
couple most strongly with the drift of streaks over spanwise distances of $O(y)$, in
agreement with the standard model of a hierarchy of wall-attached eddies \citep{town61}.

The evidence for streak accumulation is less clear, and we have shown that most of the
effect observed in \citet{toh2005interaction} and \citet{toh2018MFU} is due to their
requirement that streaks should have a streamwise velocity lower than the mean profile. When
this condition is removed, the streak density becomes fairly uniform, and its correlation with the
outer structures mostly disappears. In fact, it even changes sign far enough from the wall.
The latter also turns out to be due to the spurious modulating effect of the non-uniform wall shear
induced by the large outer scales. When this is taken into account, the correlation between
the outer flow and the streak density is uniformly positive, but mostly restricted to
structures below $y^+\approx 150$, suggesting that the co-supporting cycle is a wall-layer
effect that becomes less relevant as the Reynolds number increases. The reason is traced
to the lifetime of the streaks, which is too short to couple to higher and wider structures,
but which is probably artificially stabilised by the short simulation boxes in
\citet{toh2005interaction} (interestingly, this possibility was anticipated in that
paper, but largely forgotten afterwards).

Besides the examination and quantification of the top-down influence, the bottom-up branch
of the cycle is investigated by means of two numerical experiments that facilitate the
artificial isolation of the structures at different scales. The first one focuses on the
generation process of LSMs, starting from artificially assembled near-wall streaks in an
otherwise fluctuation-less flow. The correlation between the streak density and the outer
flow changes little before or after the LSM spontaneously appear and their size gradually
grows. Moreover, the streak accumulation history hardly influences the appearance of
large-scale components of the wall-normal velocity, implying that the streak merging does not
have a strong relation with the generation of the LSMs.

A final experiment concentrates on the preservation of existing LSMs. The large-scale
component of the velocity fluctuations in the near-wall region is filtered out once the flow is fully
established, but the outer LSMs are essentially unaffected during the regeneration of the
buffer region. Both experiments argue against the bottom-up branch of the
co-supporting hypothesis of \citet{toh2005interaction}.



\backsection[Funding]{
This work was supported by National Natural Science Foundation of China under Grant Nos.
91752205, and by the European Research Council under the Coturb and Caust grants
ERC-2014.AdG-669505 and ERC-2020.AdG-101018287.}

\backsection[Declaration of interests]{The authors report no conflict of interest.}

\backsection[Author ORCID]{Zisong Zhou, https://orcid.org/0000-0003-3708-1273; Chun-Xiao Xu, https://orcid.org/0000-0001-5292-8052; Javier Jim{\'e}nez, https://orcid.org/0000-0003-0755-843X.}

%

\bibliographystyle{jfm}
\bibliography{main}


\appendix
\section{Parameters of the PIV scheme}\label{sec:PIV}

This appendix tests the effect on the correlation coefficient defined in \r{eq:6} of the
interrogation parameters of the PIV drift estimation in \S\ref{sec:drift}.

\definecolor{purple0}{rgb}{0.63,0.13,0.94}
\definecolor{c00}{rgb}{0,1,1}
\definecolor{m00}{rgb}{1,0,1}
\definecolor{brown}{rgb}{0.62745, 0.32157, 0.17647}

\begin{figure}
  \centering
  \subfigure{
  \begin{overpic}
    [scale=0.25]{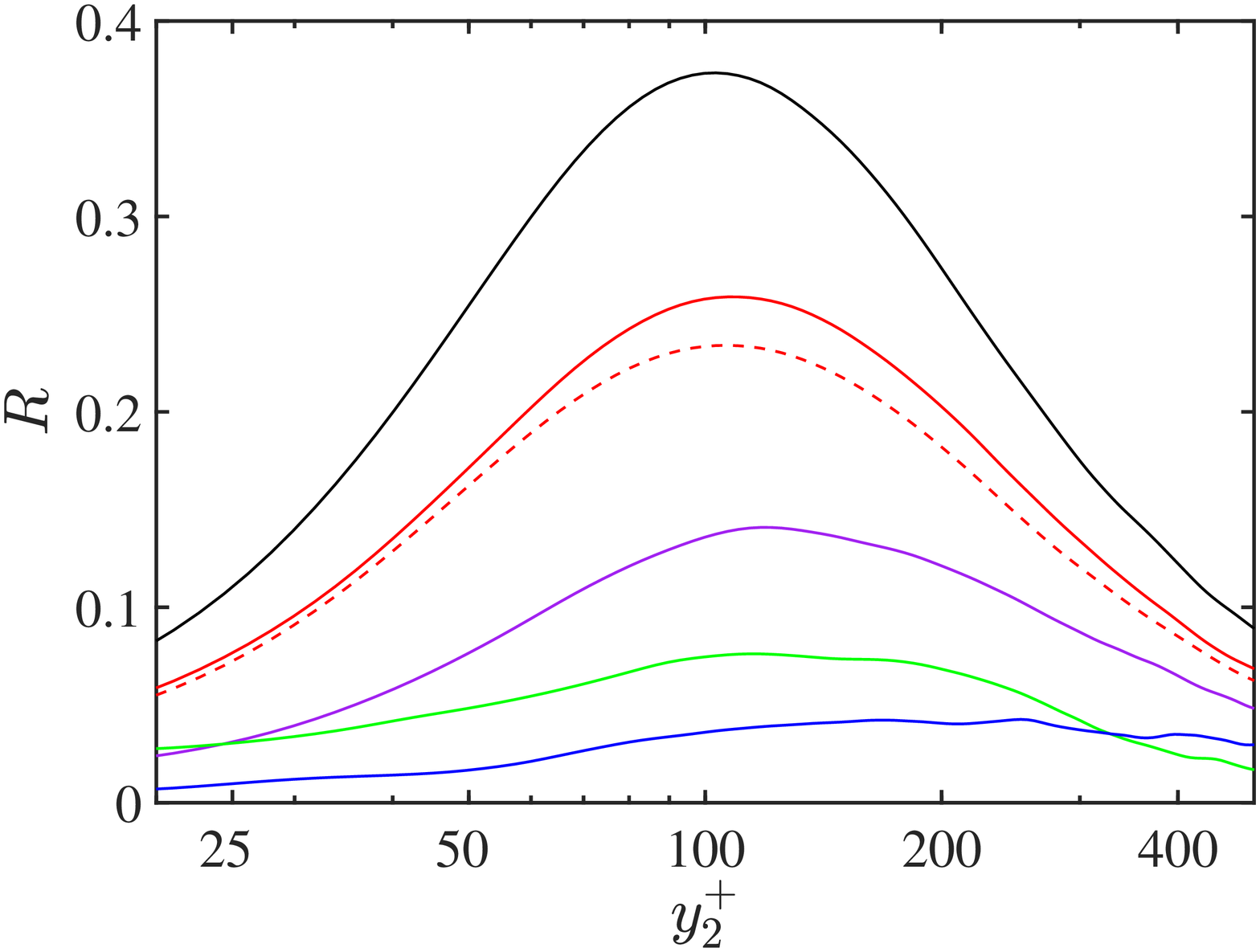}
    \put(-1,75){(\textit{a})}
  \end{overpic}
  \begin{overpic}
    [scale=0.25]{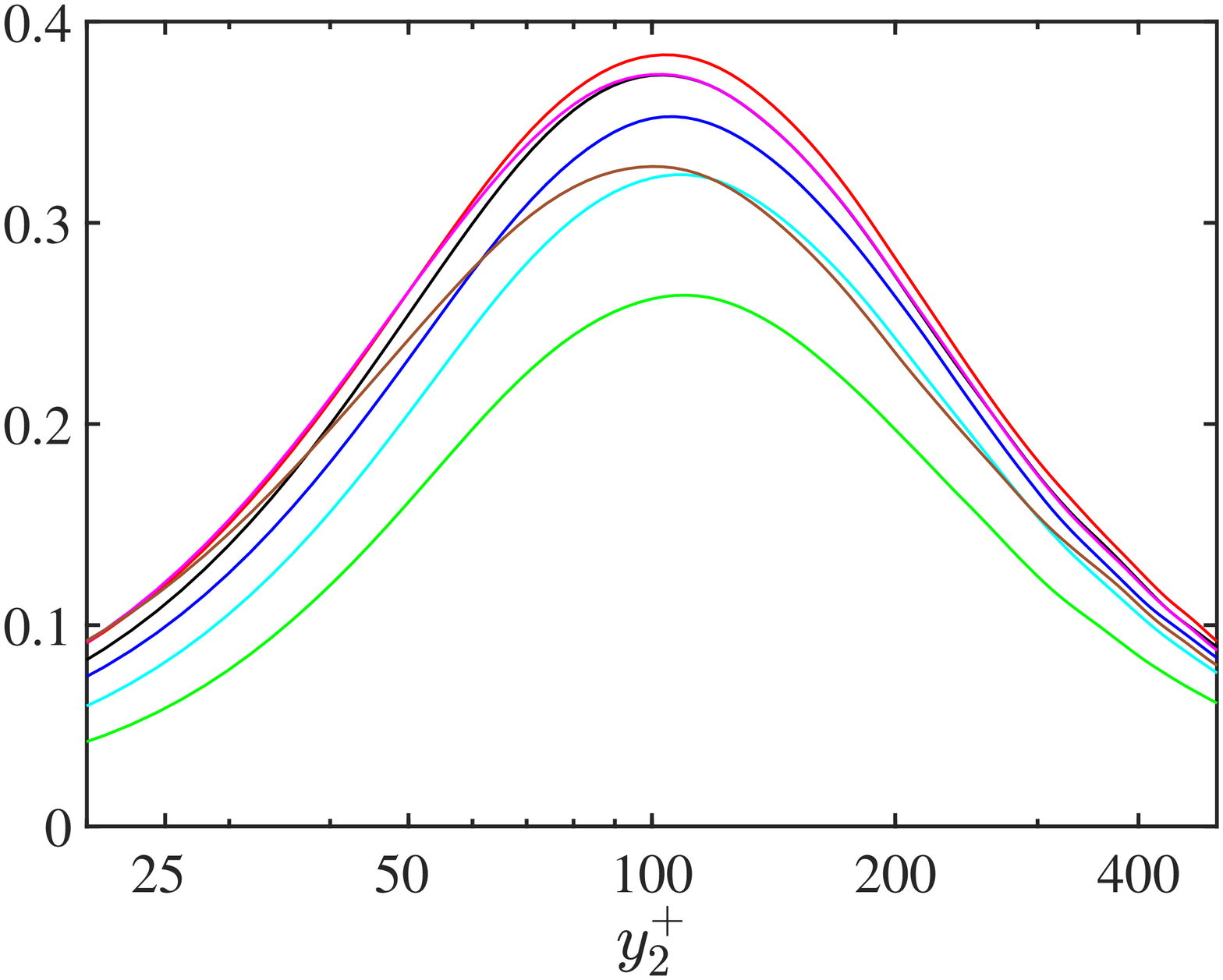}
    \put(-1,75){(\textit{b})}
  \end{overpic}
  }

  \DeclareRobustCommand\mylabela{\tikz[baseline]{\draw[solid, black, thick] (0,0.5ex) -- (0.8,0.5ex);}}
  \DeclareRobustCommand\mylabelb{\tikz[baseline]{\draw[solid, red, thick] (0,0.5ex) -- (0.8,0.5ex);}}
  \DeclareRobustCommand\mylabelc{\tikz[baseline]{\draw[solid, purple0, thick] (0,0.5ex) -- (0.8,0.5ex);}}
  \DeclareRobustCommand\mylabeld{\tikz[baseline]{\draw[solid, green, thick] (0,0.5ex) -- (0.8,0.5ex);}}
  \DeclareRobustCommand\mylabele{\tikz[baseline]{\draw[solid, blue, thick] (0,0.5ex) -- (0.8,0.5ex);}}

  \DeclareRobustCommand\mylabelf{\tikz[baseline]{\draw[solid, green, thick] (0,0.5ex) -- (0.8,0.5ex);}}
  \DeclareRobustCommand\mylabelg{\tikz[baseline]{\draw[solid, c00, thick] (0,0.5ex) -- (0.8,0.5ex);}}
  \DeclareRobustCommand\mylabelh{\tikz[baseline]{\draw[solid, blue, thick] (0,0.5ex) -- (0.8,0.5ex);}}
  \DeclareRobustCommand\mylabeli{\tikz[baseline]{\draw[solid, black, thick] (0,0.5ex) -- (0.8,0.5ex);}}
  \DeclareRobustCommand\mylabelj{\tikz[baseline]{\draw[solid, red, thick] (0,0.5ex) -- (0.8,0.5ex);}}
  \DeclareRobustCommand\mylabelk{\tikz[baseline]{\draw[solid, m00, thick] (0,0.5ex) -- (0.8,0.5ex);}}
  \DeclareRobustCommand\mylabell{\tikz[baseline]{\draw[solid, brown, thick] (0,0.5ex) -- (0.8,0.5ex);}}

  \caption{%
Correlation $R(w_s,\partial\widetilde{v}/\partial z)$, as defined in \r{eq:6}, against
$y_{2}^{+}$.
(\textit{a}) At different $\Delta t^{+}$. $u_{ad}^+=8.0$. \mylabela, $\Delta t^{+}=10.2$;
\mylabelb, $20.3$; \mylabelc, $40.6$; \mylabeld, $60.8$; \mylabele, $101.4$. The dashed red
line is drawn using only values at the streak axes.
(\textit{b}) Correlation against $y_{2}^{+}$ at different streamwise tracking speeds.
$\Delta t^{+}=20.3$. \mylabelf, $u_{ad}^+=0$; \mylabelg, $4$; \mylabelh, $6$; \mylabeli,
$8$; \mylabelj, $10$ \mylabelk, $12$; \mylabell, $16$. Both panels are for case M950 and
$\Delta z^+=214$.
}
\label{fig:RM950tutau}
\end{figure}

Figure \ref{fig:RM950tutau}(\textit{a}) displays the effect of the measurement interval $\Delta t$ on the
correlation $R({w_s,\frac{\partial\widetilde{v}}{\partial z}})$. In general, the correlation
drops as $\Delta t$ grows, and becomes too weak to be useful for $\Delta t^{+} \gtrsim 60$.
The most probable reason is that the delay acts as a filter that rejects structures with
lifetimes shorter than $\Delta t$. The lifetime of ejections in the buffer layer is
known to be approximately $30$ viscous units \citep{lozano2014time}, independently of the
Reynolds number, and this is probably the reason why the correlations in figure
\ref{fig:RM950tutau}(\textit{a}) drop rapidly above $\Delta t^{+}>30$. It is also why we choose
$\Delta t^{+}=20$ in section \ref{sec:drift}. As in the line contours in figure
\ref{fig:pdft5}, the dashed red line in figure \ref{fig:RM950tutau}(\textit{a}) is computed
using only values at the streak axes. The difference with the solid line, which uses the full
field of the drift velocity, is also minor in this case.

Figure \ref{fig:RM950tutau}(\textit{b}) shows the effect of the advection velocity of the
interrogation windows, which is relatively small except in extreme cases such as $u_{ad}=0$.
The PIV correlation does not itself use streamwise information, and the estimation of $w_s$
should work as long as the same point of the streak can be identified at the two instants used in
the correlation. Since streaks tend to be long and oriented streamwise, small advection
misalignments are not critical, but large advection errors risk mixing the irregularities of
the streak geometry, such as meandering, with its spanwise drift. The optimum correlation is obtained in
figure \ref{fig:RM950tutau}(\textit{b}) when using the advection velocity obtained elsewhere
by physical arguments ($u_{ad}^+=8$).

However, it is clear from figure \ref{fig:RM950tutau} that the largest effect on the
correlation is not from the PIV parameters, but from the choice of the correct smoothing
filter for the wall-distance of interest. As shown by figure \ref{fig:kiRet} in the body of
the paper, outer structures of a given height couple most strongly with spanwise regions of
a particular width in the buffer layer.

\section{The flow in the upper side of the numerical experiment}\label{sec:upperside}

\begin{figure}
  \centering
  \begin{overpic}
    [scale=0.48]{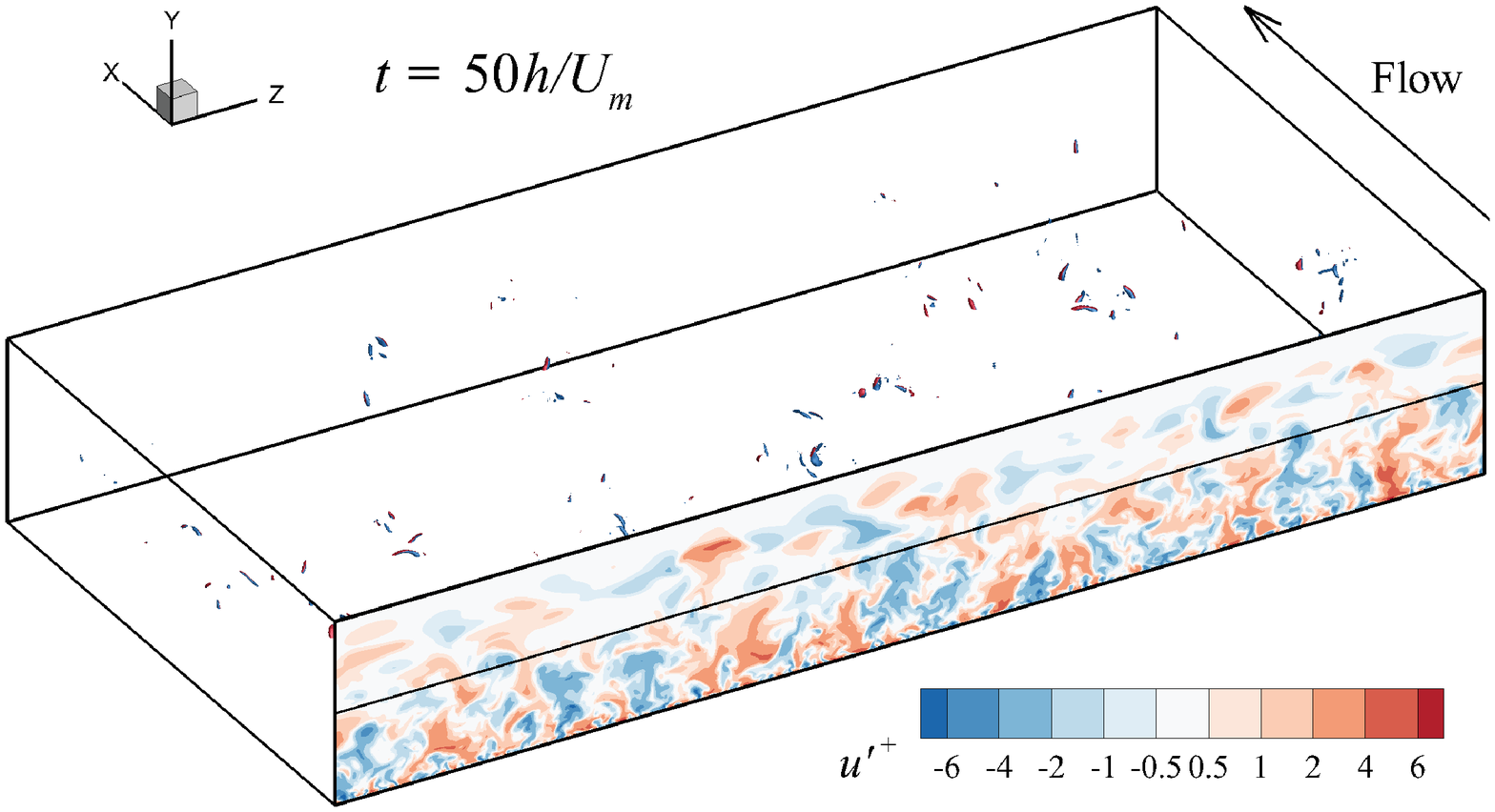}
    \put(-4,50){(\textit{a})}
  \end{overpic}

  \begin{overpic}
    [scale=0.48]{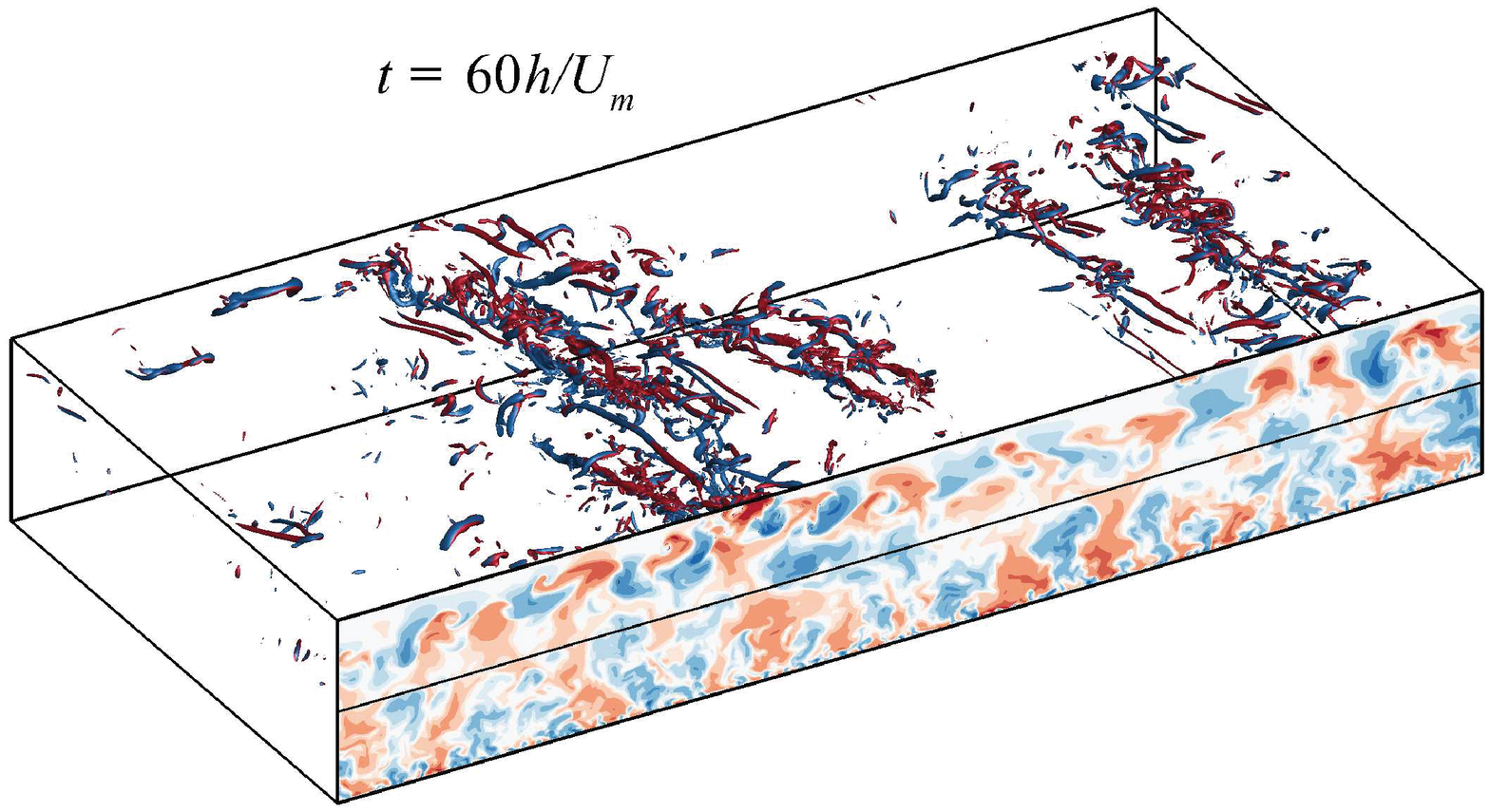}
    \put(-4,50){(\textit{b})}
  \end{overpic}

  \begin{overpic}
    [scale=0.48]{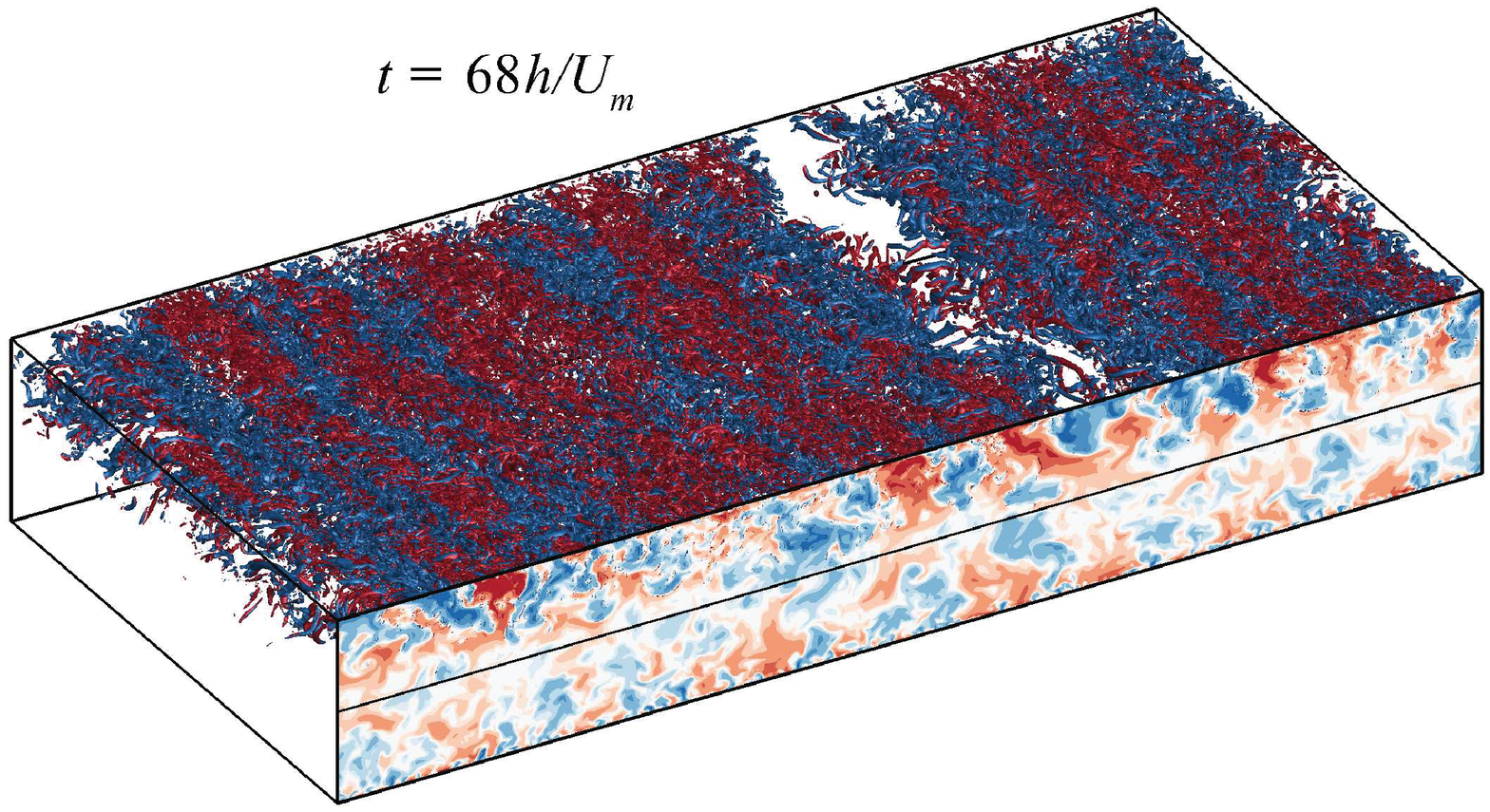}
    \put(-4,50){(\textit{c})}
  \end{overpic}

\caption{Time evolution of $u^{\prime}$ and vortices in the upper half-channel. Length in
the $x$ direction is ${2\pi}$, and in the $z$ direction is $4\pi$. The vortices are displayed
by iso-surfaces of the second invariant of the velocity gradient tensor, and coloured red
for $u^{\prime}>0$ and blue for $u^{\prime}<0$.
Flow is from bottom-right to top-left.}
\label{fig:structure2}
\end{figure}

%

This appendix discusses the flow in the upper side of the numerical experiment in \S
\ref{sec:bottomup}, where the LSMs appear before the near-wall streaks. Distributions of
vortices and streamwise velocity fluctuations before and during the transition are displayed
in figure \ref{fig:structure2}. Pairs of large-scale low- and high-speed regions form in the
upper half of the channel at $t=50$, at a time when the small-scale vortical structures are
still very rare. The LSMs gradually move closer to the wall after $t=50$, while near-wall
structures are still absent, and their intensity is progressively enhanced, with energy
drawn from the local mean shear. Small-scale vortices are first generated around $t=60$, in the strongest
high-speed regions of the LSMs, accompanied by an increase at the local
$\Rey_{\tau}$ on the upper wall (see figure \ref{fig:u_tau}). The generation of vortices
then intensifies and quickly spreads to other regions of the upper wall, whose $\Rey_{\tau}$
grows rapidly. At $t=68$, the small-scale vortical structures occupy most of the upper wall
region, and the near-wall small-scale streaks take their final shape. This evolution
suggests that the LSMs can grow and sustain themselves under the influence of the mean
velocity profile, even without near-wall streaks.

\end{document}